\documentclass[preprintnumbers]{revtex4}

\usepackage{graphicx}
\usepackage{epsf}
\usepackage{amsmath,amssymb}
\usepackage{bm}
\usepackage{color}
\usepackage{textcomp}

\newcommand{\bvec}[1]{\mbox{\boldmath $#1$}}

\begin{document}
\preprint{KUNS-2492}

\title{Approximation of reduced width amplitude and application to 
cluster decay width}

\title{Approximation of reduced width amplitude and application to 
cluster decay width}
\author{Yoshiko Kanada-En'yo}
\affiliation{Department of Physics, Kyoto University, Kyoto 606-8502, Japan}
\author{Tadahiro Suhara}
\affiliation{Matsue College of Technology, Matsue 690-8518, Japan}
\author{Yasutaka Taniguchi}
\affiliation{Center for Computational Sciences, University of Tsukuba, Tsukuba 305-8571, Japan\\
Nihon Institute of Medical Science, Moroyama-machi, Iruma-gun, Saitama 350-0435, Japan}

\begin{abstract}
We propose a simple 
method to approximately evaluate reduced width amplitude (RWA)
of a two-body spinless cluster channel 
using the norm overlap with the Brink-Bloch cluster wave function
at the channel radius.  
The applicability of the present approximation is tested for 
the $^{16}$O+$\alpha$ channel in $^{20}$Ne as well as 
the $\alpha$+$\alpha$ channel in $^8$Be.
The approximation is found to be reasonable to 
evaluate the RWA for states near the threshold energy 
and it is useful to estimate the 
$\alpha$-decay width of resonance states.
The approximation is also applied to $^9$Li, and the partial decay 
width of the $^6$He($0^+_1$)+$t$ channel is discussed. 
\end{abstract}


\maketitle
\section{Introduction}
In the recent experimental and theoretical studies, 
it has been revealed that a variety of cluster structures appear 
in various stable and unstable  nuclei in a wide mass-number region
(for instance, Refs.~\cite{Ohkubo-rev,Oertzen-rev,AMDsupp-rev,Horiuchi-rev} and references therein). 
As predicted in Ikeda's threshold rule \cite{Ikeda68,Ikeda72-supp},
remarkable cluster structures with spatial development 
have been suggested in excited states near the threshold energy. 
Interestingly, in neutron-rich nuclei, 
various cluster states containing exotic clusters have been suggested: 
He+He cluster states in Be isotopes \cite{Oertzen-rev,AMDsupp-rev,SEYA,OERTZEN,OERTZENa,ARAI,Dote:1997zz,Fujimura:1999zz,ENYObe10,ITAGAKI,OGAWA,Arai01,Descouvemont02,KanadaEn'yo:2002ay,KanadaEn'yo:2003ue,Ito:2003px,Arai:2004yf,Ito:2005yy,Pei:2006xg,Ito:2008zza,Soic96,FREER,Liendo:2002gx,SAITO04,Curtis:2004wr,Millin05,Freer:2006zz,Bohlen:2007qx,Curtis:2009zz}, 
$^{10}$Be+$\alpha$ states in $^{14}$C \cite{Soic:2003yg,oertzen04,Price:2007mm,Haigh:2008zz,Suhara:2010ww}, 
$^{14}$C+$\alpha$ states in $^{18}$O and their mirror states
\cite{Gai:1983zz,Descouvemont:1985zz,Gai:1987zz,Curtis:2002mg,Ashwood:2006sb,Yildiz:2006xc,Furutachi:2007vz,Fu:2008zzf,Johnson:2009kj,oertzen-o18}, 
$^{18}$O+$\alpha$ states in $^{22}$Ne \cite{Curtis:2002mg,Ashwood:2006sb,Yildiz:2006xc,Scholz:1972zz,Descouvemont:1988zz,Rogachev:2001ti,Goldberg:2004yk,Kimura:2007kz},
$^9$Li+$^6$He states in $^{15}$B \cite{KanadaEn'yo:2002ay}, 
$^6$He+$t$ states in $^9$Li \cite{KanadaEn'yo:2011nc}, and so on.

For direct evidence of clusters in nuclei, the cluster decay width is 
a probe to confirm the cluster structure in resonance states.
Theoretically, conventional cluster models such as the resonating group method (RGM) 
\cite{RGM,wildermuth58}
and the generator coordinate method (GCM) \cite{GCM,brink66} have been applied to study 
typical cluster structures in light stable nuclei such as the
$\alpha$+$\alpha$ structure in $^8$Be and the $^{16}$O+$\alpha$ structure in $^{20}$Ne,
and they have succeeded to describe cluster decay widths of resonance states
\cite{Ikeda72-supp,tamagaki65,wildermuth72,Nemoto72,Matsuse73,Matsuse75,Ikeda77-supp}. 

As the variation of constituent clusters becomes richer in unstable nuclei
than the well-known cluster structures in stable nuclei,
conventional cluster models based on the assumption of specific clusters 
such as $\alpha$ and $^{16}$O are no longer applicable 
for new cluster states having exotic clusters 
as  $t$, $^6$He, $^8$He, $^{10}$Be, $^{14}$C, and $^{18}$O.
For such exotic clusters, it is important to take into account 
cluster polarization, breaking, and formation as well as effects of channel coupling.
For cluster study of unstable nuclei, 
many extended frameworks  
such as antisymmetrized molecular dynamics (AMD) \cite{AMDsupp-rev,KanadaEnyo:1994kw,ENYObc,KanadaEn'yo:1998rf} 
and fermionic molecular dynamics (FMD) methods \cite{Feldmeier:1994he,Feldmeier:2000cn,Roth:2004ua,Neff:2010nm}, 
and extended cluster models of the stochastic variational method \cite{Varga:1993wp,Arai01}, the GCM method, and the generalized two-center cluster model \cite{Ito:2003px,Ito:2005yy}
have been developed.

One of the advantages of the AMD method is that the framework does not rely on the assumption of 
any clusters. Nevertheless the model wave function can describe various cluster structures as well as
one-center structures expressed by a shell-model configuration
as the formation and dissociation of clusters are automatically obtained 
in the energy variation. The method has been applied to stable and unstable nuclei and proved to be 
useful for study of cluster structures in general nuclei. 
In spite of the flexibility of the AMD wave function, 
its application to cluster decay widths is very limited.
One of the main origins for the difficulty is 
that internal wave functions of exotic clusters are generally 
more complicated than typical clusters 
which can be often expressed by a simple shell-model configuration.  
In such a case, it needs a large numerical cost to describe the details of 
asymptotic inter-cluster wave functions mainly because 
superposition of wave functions is needed in describing exotic clusters.
The FMD framework, whose wave function is quite similar to the AMD one, 
has been applied to 
a scattering problem by Neff {\it et al.}  \cite{Neff:2010nm}, but,  
the application is limited only to very light nuclei.
In many works of cluster structures with the AMD method, 
cluster resonance states are often described in a bound state approximation
and their widths are hardly discussed.

Our aim is to estimate cluster decay widths by measuring
the cluster probability at the surface
for general $A$-body wave functions containing exotic clusters or  
non-cluster components.
According to the $R$-matrix theory of nuclear reaction, the cluster decay width 
is given by the reduced width amplitude (RWA) at a channel radius
where the interaction and the antisymmetrization effect of nucleons
between clusters vanish. The method with the RWA is often used to 
estimate the cluster decay width in traditional cluster models 
within a bound-state approximation.
It means that 
if one has a reliable value of the RWA in an $A$-body wave function, 
it is able to estimate the cluster decay width 
following the RWA method as done in cluster models.
However, to extract the RWA for exotic clusters from a total wave function, 
one may encounter another problem 
because it is not obvious how to separate the partial-wave inter-cluster wave function and cluster 
internal wave functions under the antisymmetrization operator of nucleons between clusters. 
Instead, it is easier to calculate the norm overlap of the total wave function 
with the reference cluster wave functions where clusters are localized around a certain position rather
than to directly extract the RWA. 
Even for exotic clusters described by rather complicated configurations,  
the calculation of the norm overlap is usually feasible.
In the region of our interest where the effect of antisymmetrization of nucleons 
between clusters is negligible,
the norm overlap indicates the cluster probability at a certain channel radius and it should relate 
to the RWA.

In this paper, we propose a simple method to approximately calculate the RWA at the surface region 
using the norm overlap 
with the reference cluster wave function. To check the 
validity of the present approximation of the RWA, we compare the approximated RWA with the exact 
RWA in the well-known cluster states; $^{16}$O+$\alpha$ in $^{20}$Ne and 
$\alpha$+$\alpha$ states in $^8$Be in the traditional cluster-GCM calculations. 
We also make similar analysis for the $^{20}$Ne wave functions with the mixing 
of non-cluster components obtained with the AMD method. 
We show the applicability of the present approximation 
to discuss the $\alpha$-decay widths of cluster resonance states in $^{20}$Ne and $^8$Be.
As an example of application to neutron-rich nuclei, 
we apply the present method to $^9$Li and discuss the partial width of $t$ decay from
the $^6$He+$t$ cluster resonances suggested in the previous work in Ref.~\cite{KanadaEn'yo:2011nc}.

The paper is organized as follows;
In the next section, we explain the conventional cluster-GCM model
with Brink-Bloch (BB) cluster wave functions, and describe the RWA and its relation to
the decay width in the cluster model. 
The method to approximately calculate the RWA is proposed in Sec.~\ref{sec:approximation}.
The AMD framework is briefly reviewed in Sec.~\ref{sec:AMD}.
The application of the present method is demonstrated in Sec.~\ref{sec:application}, and 
finally a summary and outlooks are given in Sec.~\ref{sec:summary}.

\section{Cluster wave functions and reduced width amplitude}
\label{sec:formulation}
In this section, we review the traditional cluster-GCM model and the RWA.
For more details, the reader is refereed to the review article \cite{Ikeda77-supp}
and references therein. 

\subsection{BB cluster model and GCM wave functions}
Let us consider a system composed of two spinless clusters $C_1$ and $C_2$ with mass numbers $A_1$ and $A_2$, respectively.
In the GCM of the $C_1$+$C_2$ cluster model, the total wave function 
can be expressed by the linear combination of 
BB cluster model wave functions \cite{brink66}.

A BB cluster model wave function of the two-cluster $C_1$+$C_2$ system with the 
relative position $\bvec{S}$ is expressed as 
\begin{eqnarray}
&&|\Phi_{\rm Brink-Bloch}(\bvec{S})\rangle = |\frac{1}{\sqrt{A!}}{\cal A} \{ \psi(C_1,\frac{-A_2}{A}\bvec{S}) \psi(C_2,\frac{A_1}{A}\bvec{S})\} \rangle.
\end{eqnarray}
Here $\psi(C_i,\bvec{S}_i)$ is the wave function of the $C_i$ cluster localized around $\bvec{S}_i$,
and it is given by the harmonic oscillator (H.O.) shell model wave function with the shifted center at 
$\bvec{S}_i$. We choose the same width of H.O. for $C_1$ and $C_2$.
We set the relative position $\bvec{S}$ on the $z$-axis 
$\bvec{S}=(0,0,S)$, and for simplicity, rewrite the BB wave function 
parametrized by the inter-cluster distance $|\bvec{S}|=S$ as
\begin{equation}
|\Phi_{\rm BB}(S)\rangle \equiv 
|\Phi_{\rm Brink-Bloch}(\bvec{S}=(0,0,S))\rangle.
\end{equation}
We define the normalized $J^\pi$-projected BB wave function,
\begin{eqnarray}
|\Phi^{J\pi}_{\rm BB}(S_k) \rangle &\equiv& 
\frac{1}{\sqrt{n^l_0(S_k)}} P^{J\pi}_{00} |\Phi_{\rm BB}(S_k)\rangle,\\
P^{J\pi}_{MK}&\equiv&P^\pi P^J_{MK}\\
P^{\pi=\pm}&=&\frac{1\pm P_r}{2}\\
P^J_{MK}&=&\frac{2J+1}{8\pi^2}\int d\Omega D^{J*}_{MK}(\Omega)R(\Omega).
\end{eqnarray}
$P^\pi$ and $P^{J}_{MK}$ are the parity and total angular-momentum (spin) 
projection operators.
The BB wave function of two spinless clusters with $\bvec{S}=(0,0,S)$ is 
the $K=0$ eigen state, and its $J$-projected state
is the parity $\pi=(-1)^J$ eigen state where the inter-cluster wave function is projected onto 
the partial $l=J$ wave.
The normalization factor $n^l_0(S_k)$ is chosen to be 
$n^l_0(S_k) = \langle \Phi_{\rm BB}(S_k) |P^{J\pi}_{00}P^{J\pi}_{00} |
\Phi_{\rm BB}(S_k) \rangle$ 
so as to satisfy $|\Phi^{J\pi}_{\rm BB}(S_k)|^2=1$.

The cluster-GCM wave function for a $J^\pi$ state 
is given by the linear combination of the projected BB wave functions, 
\begin{eqnarray}\label{eq:gcm-BB}
|\Phi_{\rm GCM}\rangle &=&\sum_k c_k |\Phi^{J\pi}_{\rm BB}(S_k)\rangle.
\end{eqnarray}
Coefficients $c_k$ are determined by 
solving the discretized Hill-Wheeler equation which is equivalent to the diagonalization of
the norm and Hamiltonian matrices.
Here, the cluster-GCM wave function $\Phi_{\rm GCM}$ is normalized as
$\langle \Phi_{\rm GCM}|\Phi_{\rm GCM} \rangle=1$.
In $|\Phi_{\rm BB}(S_k)\rangle$, the relative wave function between clusters is written by a localized
Gaussian wave packet, and its partial wave expansion is given as follows.
\begin{eqnarray}
&&|\Phi_{\rm BB}(S_k)\rangle = |\frac{1}{\sqrt{A!}}{\cal A} \{ 
\Gamma(\bvec{r},\bvec{S}=(0,0,S_k),\gamma) \phi(C_1)\phi(C_2) \phi_{\rm c.m.} \}
\rangle, \\
&&\Gamma(\bvec{r},\bvec{S},\gamma)= \left ( \frac{2\gamma}{\pi}\right )^{3/4}
e^{-\gamma(\bvec{r}-\bvec{S})^2} =\sum_l\Gamma_l(r,S,\gamma)\sum_m 
Y_{lm}(\hat{\bvec{r}})Y^*_{lm}(\hat{\bvec{S}}),\\
&& \Gamma_l(r,S,\gamma) \equiv 4\pi (\frac{2\gamma}{\pi})^{\frac{3}{4}} i_l(2\gamma S r) e^{-\gamma(r^2+S^2)},\\
&& \gamma \equiv  \frac{A_1A_2}{A} \nu,\\
&&\phi_{\rm c.m.}=\left( \frac{2A\nu}{\pi} \right) e^{-A\nu\bvec{r}_G^2},
\end{eqnarray}
where $i_l$ is the modified spherical Bessel function,
$\bvec{r}$ is the relative coordinate between centers of mass of clusters, 
$\phi(C_1)$ and $\phi(C_2)$ are internal wave functions of clusters
$\bvec{r}_G$ is the center of mass coordinate and 
$\phi_{\rm c.m.}$ is the wave function of the total center of mass motion (c.m.m.).
$\nu$ is the width parameter for the H.O. for clusters, and relates to the width $b$ of the H.O. as
$\nu\equiv 1/2b^2$.
Then, in the projected BB 
wave function $|\Phi^{J\pi}_{\rm BB}(S_k)\rangle$, 
the radial part $\chi^{\rm BB}_l(S_k;r)$ of the $l$-wave relative 
wave function is written with the function $\Gamma_l$;
\begin{eqnarray} \label{eq:chi}
&& |\Phi^{J\pi}_{\rm BB}(S_k)\rangle= 
 |\frac{1}{\sqrt{A!}}{\cal A} \{ \chi^{\rm BB}_l(S_k;r)
Y_{l0}(\hat{r}) \phi(C_1)\phi(C_2) \phi_{\rm cm} \}\rangle, \\
&& \chi^{\rm BB}_l(S_k;r)=\frac{1}{\sqrt{n^l_0(S_k)}} \sqrt{\frac{2l+1}{4\pi}}\Gamma_l(r,S_k,\gamma).
\end{eqnarray}
Here the relation $Y^*_{l0}(\hat{\bvec{S}})=\sqrt{\frac{2l+1}{4\pi}}$ for 
$\bvec{S}=(0,0,S_k)$ is used.

Using $\chi^{\rm BB}_l(S_k;r)$, 
the cluster-GCM wave function is also 
rewritten in the form consisting of the relative wave function, 
internal wave functions of clusters, and the c.m. wave function,
\begin{eqnarray}
&&\Phi_{\rm GCM}=\sum_k c_k |\Phi^{J\pi}_{\rm BB}(S_k)\rangle =| \frac{1}{\sqrt{A!}} \mathcal{A} \left[ \chi^{\rm GCM}_l(r) Y_{l0}(\hat{r}) \phi(C_1)\phi(C_2)\phi_{\rm c.m.}  \right]  \rangle \\
&&\chi^{\rm GCM}_l(r)=\sum_k c_k  \chi^{\rm BB}_l(S_k;r) \sum_k c_k  \sqrt{\frac{2l+1}{4\pi}} 
\Gamma_l(r,S_k,\gamma).
\end{eqnarray}

\subsection{Reduced width amplitude}
For a wave function $\Psi$ of the $A$-nucleon system, 
the RWA $ry_l(r)$ 
for the $C_1$+$C_2$ cluster channel is defined as
\begin{eqnarray}
ry_l(r)\equiv r \sqrt{\frac{A!}{A_1!A_2!}} \langle Y_{l0}(\hat{r}) \phi(C_1)\phi(C_2) |\Psi\rangle. 
\end{eqnarray}
Here $\Psi$ does not contain the c.m.m. 
$y_l(r)$ is regarded as the radial part of a relative wave function
where the antisymmetrization effect is taken into account.

For a RGM-type cluster wave function of the $C_1$+$C_2$ system, 
\begin{eqnarray}
\Psi = \frac{1}{\sqrt{A!}} \mathcal{A} \left[ \chi_l(r) Y_{l0}(\hat{r}) \phi(C_1)\phi(C_2) \right]  \rangle,
\end{eqnarray}
$y_l$ is calculated by using the expansion of $\chi_l(r)$ 
with the orthonormal set $R_{nl}(r)$ of the radial wave functions 
of H.O. with the width parameter 
$b=1/\sqrt{2\gamma}$  given by $\gamma=\nu A_1 A_2/A$,
\begin{eqnarray}
\chi_l(r)&=&\sum_n a_n R_{nl}(r),\\
a_n&=& \int r^2dr R_{nl}(r) \chi_l(r),\\
y_l(r)&=&\sum_n a_n \mu_{nl} R_{nl}(r),
\end{eqnarray}
$\mu_{nl}$ is the eigen value of the RGM norm kernel \cite{Ikeda77-supp}. 
We also define the function $u(r)$ as 
\begin{eqnarray}
u_l(r)&=&\sum_n  a_n \sqrt{\mu_{nl}} R_{nl}(r).
\end{eqnarray}
For the normalized cluster wave function $\langle \Psi|\Psi \rangle=1$, 
the function $u_l(r)$ also satisfies the normalization,
\begin{eqnarray}
\int |u_l(r)|^2  r^2 dr=1.
\end{eqnarray}
The spectroscopic factor $S$ is defined by the RWA,
\begin{eqnarray}
S = \int |y_l(r)|^2  r^2 dr.
\end{eqnarray}
Functions, $\chi_l(r)$, $y_l(r)$, and $u_l(r)$ are interpreted as 
inter-cluster wave functions, {\it i.e.}, 
the radial part of relative wave functions, but they are different 
in the treatment of the antisymmetrization effect between clusters. 
$\chi_l(r)$ is the relative wave function before antisymmetrization
and can contain unphysical forbidden states with $\mu_{nl}=0$. 
In the functions, $y_l(r)$ and $u_l(r)$, the antisymmetrization is taken
into account and all forbidden states are excluded in both functions,
but the treatment of partially allowed states with $\mu_{nl}\ne 1$ is
different.
All the functions $\chi_l(r)$, $y_l(r)$, and $u_l(r)$ have 
the same asymptotic behavior in the large $r$ region where 
the antisymmetrization effect between clusters vanishes while they are different 
in the inner region where clusters largely overlap with each other 
and feel the strong antisymmetrization effect.

\subsection{Inter-cluster wave functions for GCM and BB wave functions}
For the cluster-GCM wave function, the inter-cluster wave functions $y^{\rm GCM}_l(r)$ 
is calculated from the function $\chi^{\rm GCM}_l(r)$, 
\begin{eqnarray}
\chi^{\rm GCM}_l(r)&=&\sum_n a_n R_{nl}(r),\\
a_n&=& \int r^2 dr R_{nl}(r) \chi^{\rm GCM}_l(r),\\
y^{\rm GCM}_l(r)&=&\sum_n a_n \mu_{nl} R_{nl}(r).
\end{eqnarray}
$ry^{\rm GCM}_l(r)$ is the RWA of the cluster-GCM wave function.

Also for the $J^\pi$-projected BB wave function $\Phi^{J\pi}_{\rm BB}(S_k)$, 
we can define the antisymmetrized inter-cluster wave functions 
$y^{\rm BB}_l(r)$ and $u^{\rm BB}_l(r)$ from the non-antisymmetrized 
wave function $\chi^{\rm BB}_l(r)$,  
\begin{eqnarray}
y^{\rm BB}_{l}(S_k;r)= \sum_{n} a_n \mu_{nl} R_{nl}(r),\\
u^{\rm BB}_{l}(S_k;r)= \sum_{n} a_n \sqrt{\mu_{nl}} R_{nl}(r),\\
a_n= \int r^2 dr R_{nl}(r) \chi^{\rm BB}_{l}(S_k;r).
\end{eqnarray}
Here, the normalization $\int |u^{\rm BB}_{l}(S_k;r)|^2 r^2 dr=1$ is satisfied because of the condition $|\Phi^{J\pi}_{\rm BB}(S_k)|^2=1$.

\subsection{Antisymmetrization effect between clusters}
A BB wave function is parametrized by the inter-cluster distance parameter
$S_k$. The non-antisymmetrized wave function
$\chi^{\rm BB}_l(S_k;r)$ is the function localized around $r=S_k$. 
In case of a small $S_k$, clusters largely overlap with each other and 
the inter-cluster wave function is strongly affected by the antisymmetrization of nucleons
between clusters. For such a small $S_k$, 
$\chi^{\rm BB}_l(S_k;r)$ contains much component of unphysical forbidden states,
which do not affect the total wave function $\Phi^{J\pi}_{\rm BB}(S_k)$.
Since the forbidden states are excluded in $u^{\rm BB}_l(r)$ as well as $y^{\rm BB}_l(r)$, 
the norm of the original function $\chi^{\rm BB}_l(S_k;r)$ is usually larger than 
that of $u^{\rm BB}_l(r)$. In other words, 
because of the antisymmetrization the norm of inter-cluster 
wave function $u^{\rm BB}_l(r)$ is relatively small 
compared with the original one $\chi^{\rm BB}_l(S_k;r)$. 
We call the ratio of norms
\begin{eqnarray}
{\cal N}_l(S_k) \equiv \frac{\int |u^{\rm BB}_l(r)|^2 r^2 dr}{\int 
|\chi^{\rm BB}_l(S_k;r)|^2 r^2 dr}
= \frac{1}{\int |\chi^{\rm BB}_l(S_k;r)|^2 r^2 dr},
\end{eqnarray}
``allowedness factor'' which indicates the weakness of 
the antisymmetrization effect between clusters. 
In a BB wave function with an enough 
large $S_k$ where the antisymmetrization effect between clusters 
is negligible,  ${\cal N}_l(S_k)\simeq 1$.
In such a case, the antisymmetrized inter-cluster wave functions 
$y^{\rm BB}_{l}(S_k;r)$ and $u^{\rm BB}_{l}(S_k;r)$ are consistent with the original non-antisymmetrized wave function $\chi^{\rm BB}_{l}(S_k;r)$.
In a small $S_k$ limit, 
$\chi^{\rm BB}_{l}(S_k;r)$ is dominated by unphysical forbidden states and 
the ratio of the norms for the physical inter-cluster wave funtion 
$\int |u^{\rm BB}_l(r)|^2 r^2 dr$ to that for 
$\int|\chi^{\rm BB}_l(S_k;r)|^2 r^2 dr$ goes to zero, 
{\it i.e.}, the allowedness factor ${\cal N}_l(S_k)\approx 0$ indicating the 
strong limit of the antisymmetrization effect.

Moreover, the relative wave function $\chi^{\rm BB}_{l}(S_k;r)$ 
has a peak structure around $r=S_k$. 
It means that, for a large $S_k$, 
the function $r \chi^{\rm BB}_{l}(S_k;r)\approx r y^{\rm BB}_{l}(S_k;r)
\approx r u^{\rm BB}_{l}(S_k;r)$ is a localized function 
around $r=S_k$.

\subsection{Decay width and RWA}
For the $\alpha$-decay width $\Gamma_\alpha$, the reduced width $\gamma^2_\alpha(a)$ at the channel radius $a$ is defined
\begin{eqnarray}
\Gamma_\alpha=2P_l(a)\gamma^2_\alpha(a),\\
P_l(a)=\frac{ka}{F^2_l(ka)+G^2_l(ka)},
\end{eqnarray}
where $F_l$ and $G_l$ are the regular and irregular Coulomb functions, respectively, 
$k$ is the momentum of inter-cluster motion in the asymptotic region, and 
$mu$ is the reduced mass.
A dimensionless reduced $\alpha$-width $\theta^2_\alpha(a)$ defined by the
ratio of the reduced $\alpha$-width $\gamma^2_\alpha(a)$ to its Wigner limit 
$\gamma^2_W(a)=3\hbar^2/2\mu a^2$, 
\begin{eqnarray}
\theta^2_\alpha(a)=\gamma^2_\alpha(a)/\gamma^2_W(a),
\end{eqnarray}
is a good measure to discuss the $\alpha$-cluster probability at the surface.
The experimental value of $\theta^2_\alpha(a)$ is deduced from  
the measured $\alpha$-decay width $\Gamma_\alpha$ of the resonance states.

On the other hand, according to the $R$-matrix theory of nuclear reaction, the reduced width is approximately given by the RWA $ay_l(a)$ for the $\alpha$ cluster channel, 
\begin{eqnarray}\label{eq:rw-app}
\gamma^2_\alpha(a)&=&\frac{\hbar^2}{2\mu a}\left[ ay_l(a) \right]^2\\
\theta^2_\alpha(a)&=&\frac{a}{3}\left[ ay_l(a) \right]^2.
\end{eqnarray}
This approximation is good especially for narrow resonances.
In the theoretical calculation using a bound state approximation, 
the above $R$-matrix based approximation is often used to estimate the width from the 
calculated RWA $ay_l(a)$.

\section{Approximated RWA}\label{sec:approximation}

As mentioned above, the partial 
decay width can be estimated using the RWA $ry_l(r)$ at a channel radius $r=a$
with the relation given in Eq.~(\ref{eq:rw-app}) based on the $R$-matrix theory.
Our aim here is to approximately evaluate the RWA $ry_l(r)$ at a certain 
channel radius for cluster-GCM wave functions $\Phi_{\rm GCM}$ or 
more general $A$-body wave functions $\Phi$ having cluster 
breaking components in order 
to estimate the partial width of cluster decay from resonance states.
In general, clusters, $C_1$ and $C_2$, are not shell-closed clusters and they have
more complicated configurations than shell-closed nuclei.
If clusters are deformed and their intrinsic wave functions 
are not spin-parity eigen wave functions, spin-parity 
projections of subsystems (clusters) are needed to calculate exact RWA
and it usually enlarges the numerical cost.
Moreover, it is not necessarily easy 
to solve the eigen value problem of the RGM norm kernel 
for the $C_1$+$C_2$ channel except for the case of simple clusters. 

Alternatively, we propose a method to calculate an approximated value of 
the RWA $ay_l(a)$ using the simple overlap norm of $\Phi$ with a 
single BB cluster wave function parametrized by $S_k=a$.
The region for the channel radius $a$ of our interest
is the surface region 
where the inter-cluster distance is large enough to ignore
the antisymmetrization effect of nucleons between clusters. 
Let us consider the projected BB wave function $\Phi^{J\pi}_{\rm BB}(S_k)$
with $S_k=a$ which is localized around the channel radius $a$. 
The overlap of $\Phi$ with $\Phi^{J\pi}_{\rm BB}(S_k)$ can be calculated 
as the norm overlap of antisymmetrized $A$-body wave functions. It is also 
given by the overlap of the inter-cluster wave functions $u_l(r)$
for $\Phi$ and $u^{\rm BB}_l(r)$ for $\Phi^{J\pi}_{\rm BB}$, 
\begin{eqnarray}\label{eq:overlap}
|\langle \Phi | \Phi^{J\pi}_{\rm BB}(S_k) \rangle |=| \langle ru_l(r)| 
ru^{\rm BB}_{l}(S_k;r) \rangle |.
\end{eqnarray}
Here we define
\begin{eqnarray}
\langle f(r)|g(r) \rangle\equiv \int^\infty_0 f^*(r)g(r) dr. 
\end{eqnarray}
As mentioned before, in the region around $a$ 
where the antisymmetrization effect between clusters is negligible, 
$\chi(r)\approx y_l(r)\approx u_l(r)$ for $\Phi$, and also
$\chi^{\rm BB}_l (S_k;r)\approx y^{\rm BB}_{l}(S_k;r) 
\approx u^{\rm BB}_{l}(S_k;r)$ for $\Phi^{J\pi}_{\rm BB}(S_k)$.
Moreover, for simplicity, we approximate 
the inter-cluster wave function 
$\chi^{\rm BB}_{l}(S_k;r)$ for $\Phi^{J\pi}_{\rm BB}(S_k)$ with a 
Gaussian form, 
\begin{eqnarray}\label{eq:XG}
r \chi^{\rm BB}_{l}(S_k;r)\approx \left(\frac{2\gamma}{\pi}\right)^{1/4}
e^{-\gamma(r-S_k)^2}\equiv {\cal X}^G(S_k;r). 
\end{eqnarray}
Namely, the inter-cluster wave function for $\Phi^{J\pi}_{\rm BB}(S_k)$ is localized around $r=S_k$ with the width $2\sqrt{\gamma}$.
Let us consider here to measure the unknown RWA $ry_l(r)\approx ru_l(r)$ 
with the localized reference function 
${\cal X}^G(S_k;r)\approx ru^{\rm BB}_{l}(S_k;r)$ using the equation (\ref{eq:overlap}), 
\begin{eqnarray}
&&|\langle \Phi | \Phi^{J\pi}_{\rm BB}(S_k) \rangle |=| \langle ru_l(r)| ru^{\rm BB}_{l}(S_k;r) \rangle |
\approx  \langle ry_l(r)| {\cal X}^G(S_k;r) \rangle.
\end{eqnarray}
We assume that the RWA $ry_l(r)$ for the realistic wave function 
$\Phi$ is a gradually changing function 
compared with the localized reference function 
${\cal X}^G(S_k=a;r)$ and
it can be approximated to be constant $ay(a)$ at least 
in the region around $S_k=a$ with the width $2\sqrt{\gamma}$ where 
${\cal X}^G(S_k=a;r)$ gives a finite contribution to the integrated value
$\langle ry_l(r)| {\cal X}^G(S_k;r) \rangle$. 
In this assumption, 
the overlap is approximately given as 
\begin{eqnarray}
&&\langle ry_l(r)| {\cal X}^G(S_k=a;r) \rangle \approx ay(a) \int {\cal X}^G(S_k=a;r) dr =   ay(a) 
\sqrt{2}\left(\frac{2\gamma}{\pi}\right)^{-1/4}.
\end{eqnarray}
It means that the norm overlap with $\Phi^{J\pi}_{\rm BB}(S_k=a)$ relates to the RWA $ay_l(a)$
and we obtain the following approximation for the RWA 
\begin{eqnarray}\label{eq:y-app}
|ay_l(a)| \approx
\frac{1}{\sqrt{2}}\left(\frac{2\gamma}{\pi}\right)^{1/4}
|\langle \Phi | \Phi^{J\pi}_{\rm BB}(S_k=a)\rangle| 
\equiv ay^{\rm app}(a).
\end{eqnarray}
This approximation works reasonably for the tail part of the RWA of 
cluster states near the threshold energy because the 
inter-cluster wave function has an asymptotic tail 
determined by the energy measured from the
threshold. 
If $ry_l(r)$ is a rapidly changing function, 
the approximated function $ry^{\rm app}(r)$ 
corresponds to a smeared function 
with the resolution $2\sqrt{\gamma}$
and $ay^{\rm app}(a)$ indicates the mean value of $ry_l(r)$ around $r=a$. 
Moreover, the approximation is not valid in the small $a$ region with 
the strong antisymmetrization effect. 
However, for the present aim to 
estimate decay width of resonances using the approximated RWA, 
we can reasonably approximate the RWA with the present method
as shown later.  

\section{AMD method}\label{sec:AMD}
The AMD method is useful to describe the formation and breaking of clusters
as well as shell-model states with non-cluster structure. 
The applicability of the AMD method to stable and unstable nuclei have been proved, 
for example, in Refs.~\cite{AMDsupp-rev,KanadaEn'yo:1998rf}.
For the detailed formulation of the AMD, the reader is referred to those references.

\subsection{Formulation of AMD(VAP)}

An AMD wave function of an $A$-nucleon system is given by a Slater determinant
of Gaussian wave packets;
\begin{eqnarray}
 \Phi_{\rm AMD}({\bvec{Z}}) &=& \frac{1}{\sqrt{A!}} {\cal{A}} \{
  \varphi_1,\varphi_2,...,\varphi_A \},\\
 \varphi_i&=& \phi_{{\bvec{X}}_i}\sigma_i\tau_i,\\
 \phi_{{\bvec{X}}_i}({\bvec{r}}_j) & = &  \left(\frac{2\nu}{\pi}\right)^{4/3}
\exp\bigl\{-\nu({\bvec{r}}_j-\frac{{\bvec{X}}_i}{\sqrt{\nu}})^2\bigr\},
\label{eq:spatial}\\
 \sigma_i &=& (\frac{1}{2}+\xi_i)\sigma_{\uparrow}
 + (\frac{1}{2}-\xi_i)\sigma_{\downarrow}.
\end{eqnarray}
$\phi_{{\bvec{X}}_i}$ and $\sigma_i$ are spatial and spin functions
of the $i$th single-particle wave function, and 
$\tau_i$ is the isospin
function fixed to be up (proton) or down (neutron). 
Accordingly, an AMD wave function
is expressed by a set of variational parameters, ${\bvec{Z}}\equiv 
\{{\bvec{X}}_1,{\bvec{X}}_2,\cdots, {\bvec{X}}_A,\xi_1,\xi_2,\cdots,\xi_A \}$.
The width parameter $\nu$ is chosen to be a common value for all nucleons. 

The energy variation after spin and parity projections (VAP) 
is performed to get the AMD wave function for the lowest $J^\pi$ state.
The parameters ${\bvec{X}}_i$ and $\xi_{i}$ ($i=1\sim A$) are varied to
minimize the expectation value of the Hamiltonian,
$\langle \Phi|H|\Phi\rangle/\langle \Phi|\Phi\rangle$,
with respect to the spin-parity eigen wave function projected 
from an AMD wave function; $\Phi=P^{J\pi}_{MK}\Phi_{\rm AMD}({\bvec{Z}})$.

In the AMD model space, all single-nucleon wave functions are treated as 
independent Gaussian wave packets, and cluster formation and breaking 
are described by spacial configurations of Gaussian centers, $\bvec{X}_i$. 
If we choose a specific set of the parameters $\{\bvec{Z}\}$, 
the AMD wave function can be equivalent to a BB wave function.
For instance, the $\alpha$+$\alpha$ BB wave function with $\bvec{S}_k$ 
is expressed by the AMD wave function by taking 
$\bvec{X}_1=\cdots=\bvec{X}_4=\bvec{S}_k/2$ and 
$\bvec{X}_5=\cdots=\bvec{X}_8=-\bvec{S}_k/2$ for spin-up and down protons and neutrons. 
Similarly, it is also able to express a $^{16}$O+$\alpha$ BB wave function with 
an AMD wave function.

\subsection{Hybrid model of AMD(VAP)+cluster}

In a single AMD wave function, which is based on a single Slater determinant, 
the inter-cluster wave function does not have the correct asymptotic behavior.
However, in a realistic cluster state near the threshold energy, 
the inter-cluster wave function should have an outer tail whose asymptotic behavior is 
determined by the $\alpha$-decay energy.
To describe the detailed behavior of the outer tail, 
we perform the hybrid calculation by 
superposing the AMD(VAP) wave functions and 
$^{16}$O+$\alpha$ cluster BB wave functions as done in $^{16}$O+$^{16}$O cluster states in 
$^{32}$S by Kimura {\it et al.} \cite{Kimura:2003ue}. 

In the hybrid calculation, the wave function for the $J^\pi$ state 
is written by superposing 
the AMD wave functions $\Phi_{\rm AMD}({\bf Z}^{J'\pi})$
obtained by VAP for various $J'^{\pi}$ states and the BB wave functions,
\begin{equation}\label{eq:hybrid}
|\Phi \rangle=\sum_{J'} c(J') 
|P^{J\pi}_{MK}\Phi_{\rm AMD}({\bf Z}^{J'\pi})\rangle+ 
\sum_k c(k) |\Phi^{J\pi}_{\rm BB}(S_k)\rangle,
\end{equation}
where the coefficients $c(J')$ and $c(k)$ are determined by the
diagonalization of the norm and Hamiltonian matrices. $K=0$ is chosen in the present 
calculation of $^{20}$Ne.

\subsection{Projection to cluster model space}
For a general microscopic $A$-body wave function $\Phi$
of a spin and parity $J^\pi$ eigen state, 
we can extract the cluster components
of $\Phi$ when the c.m.m. of $\Phi$ is separable.
The AMD wave function satisfies 
this condition. 

From a set of the $J^\pi$-projected BB wave functions, 
$|\Phi^{J\pi}_{\rm BB}(S_k)\rangle$ $(k=1,\cdots,k_{\rm max})$, an
orthonormal set of wave functions, 
$|\Phi^{\rm cluster}_m\rangle$ $(m=1,\cdots,k_{\rm max})$,
is constructed.
Here $\Phi^{\rm cluster}_m$ is given by the linear combination of
the basis wave functions $\Phi^{J\pi}_{\rm BB}(S_k)$ so as to satisfy
the orthonormality
$\langle \Phi^{\rm cluster}_m|\Phi^{\rm cluster}_n\rangle=\delta_{mn}$.
By using this orthonormal set of cluster wave functions, 
the projection operator $P^{\rm cluster}$ on to the model space of 
cluster wave functions is defined as 
\begin{equation}
P^{\rm cluster}\equiv \sum_{m}  |\Phi^{\rm cluster}_m\rangle \langle\Phi^{\rm cluster}_m|
\end{equation}
The cluster component in a general wave function $\Phi$ is given by 
the expectation value of the projection operator
\begin{equation}
{\cal P}^{\rm cluster}\equiv \langle \Phi 
|P^{\rm cluster}| \Phi \rangle.
\end{equation}
For a single-channel cluster-GCM wave function, the cluster component 
${\cal P}^{\rm cluster}=1$, while if $\Phi$ contains non-cluster components 
it is smaller than 1. 
The inter-cluster wave functions
$\chi_l(r)$, $y_l(r)$ and $u_l(r)$ for the general wave function $\Phi$ 
can be calculated by projecting it onto the cluster model space expressed by
the linear combination of the BB wave functions.
For a normalized wave function $\Phi$, 
the cluster component ${\cal P}^{\rm cluster}$ can be also given 
in terms of the norm of the inter-cluster wave function $u_l(r)$ as
\begin{equation}
{\cal P}^{\rm cluster}=\langle ru_l(r) | ru_l(r) \rangle= \int |u_l(r)| r^2 dr.
\end{equation}

\section{Application of the approximated RWA}\label{sec:application}

We check the validity of the approximated 
RWA defined in (\ref{eq:y-app}) for
$^{16}$O+$\alpha$ and $\alpha$+$\alpha$ systems 
by comparing the approximated RWA with the exact value. We then apply the 
present method to $^9$Li and discuss the partial decay width 
of the $^6$He($0^+_1$)+$t$ channel for excited states of $^9$Li. 

\subsection{RWA in $^{20}$Ne} 
In $^{20}$Ne, the ground band ($K^\pi=0^+_1$), the $K^\pi=0^-$ band, and the higher-nodal 
$K^\pi=0^+$ band starting from the $J^\pi=0^+_4$ state are 
considered to be $^{16}$O+$\alpha$ cluster states because 
they are described well with $^{16}$O+$\alpha$ cluster models
except for the energy position of the $8^+$ state in the ground band. 
For the $^{16}$O+$\alpha$ cluster states, 
it is rather easy to calculate the exact RWA using the eigen values $\mu_{nl}$ 
of the RGM norm kernel because both clusters are shell-closed nuclei 
and their wave functions are given by simple H.O. configurations.

As the first test to check the present method of the approximated RWA, we 
calculate the approximated values $ry^{\rm app}_l(r)$ 
for the $^{16}$O+$\alpha$ cluster-GCM wave function
and compare them with the exact RWA.
We obtain the wave function $\Phi^{\rm GCM}$ 
for the ground and excited states of $^{20}$Ne 
with the $^{16}$O+$\alpha$ cluster-GCM calculation. 
The adopted effective interaction is Volkov No.2 with $m=0.62$ \cite{VOLKOV}. 
The width parameter $\nu=0.16$ fm$^{-2}$ is used for both $^{16}$O and $\alpha$ clusters. 
Those interaction parameters and the width parameter are the same as those used in the preceding 
study of $^{20}$Ne with the RGM by Matsuse {\it et al.} \cite{Matsuse75}. The parameter set
reproduces well the ground-band energy spectra measured 
from the threshold energy as well as the root-mean-square radius 
 of $^{16}$O. As the basis wave functions of the cluster-GCM calculation, 
ten BB wave functions with the $^{16}$O-$\alpha$ distance 
$S_k=1,2,\cdots 10$ fm are adopted. It corresponds to 
a bound state approximation. 

As an another test, we also do the similar analysis of the
RWA using AMD wave functions of $^{20}$Ne.
It is a test to check the applicability of the method for the case that 
the system is not a pure cluster state because
the AMD wave function can contain
non-cluster components as well as the cluster component. 
We perform the AMD(VAP) calculation 
to obtain the optimum solution of the AMD wave functions for the $J^\pi=0^+$, $2^+$, $\cdots$,
$8^+$ states in the ground band of $^{20}$Ne. 
As for the effective interaction,  
Volkov No.2 with $m=0.66$ supplemented by the spin-orbit force of the 
G3RS \cite{LS} with the strength $u_I=-u_{II}=2400$ MeV
is chosen so as to reproduce 
the ground band spectra measured from the threshold energy of 
the H.O. shell-closed $^{16}$O and $\alpha$ clusters. In the AMD(VAP) calculation, 
the larger Majorana parameter $m$ than that used in the cluster-GCM calculation is needed to avoid the overbinding problem 
because the extra energy is gained by the spin-orbit interaction and the cluster dissociation in the AMD(VAP) calculation.
We also perform the hybrid calculation of AMD(VAP)+cluster 
by superposing AMD(VAP) wave functions 
and $^{16}$O+$\alpha$ cluster BB wave functions using the same interaction. 

\begin{figure}[tb]
\begin{center}
	\includegraphics[width=6.5cm]{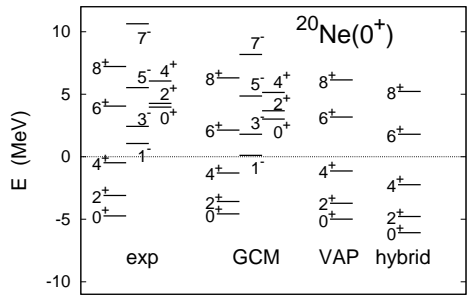} 	
\end{center}
\vspace{0.5cm}
  \caption{Energies of the ground and excited states of $^{20}$Ne
obtained with the cluster-GCM, AMD(VAP), and the hybrid(AMD+cluster) calculations.
The energies measured from the $^{16}$O+$\alpha$ threshold energy are compared with the experimental data \cite{Tilley98}.
The adopted effective interaction is Volkov No.2 with $m=0.62$ for 
for cluster-GCM, and that with $m=0.66$ supplemented by the spin-orbit term of the 
G3RS force with the strength $u_I=-u_{II}=2400$ MeV for
the AMD(VAP) and the hybrid calculations.
\label{fig:20Ne-ene}}
\end{figure}

The calculated energy levels measured from the $^{16}$O+$\alpha$ threshold
are shown in Fig.~\ref{fig:20Ne-ene} compared with the experimental energy levels
of the ground, the $K^\pi=0^-$, and the higher-nodal(hn) $K^\pi=0^+$ bands.
The $J^\pi=0^+_2$, $2^+_2$, and $4^+_2$ states obtained with the cluster-GCM calculation
correspond to the higher-nodal band members, $0^+_{\rm hn}$, $2^+_{\rm hn}$, and $4^+_{\rm hn}$ starting from the $0^+_4$ state in the experimental data.
It should be commented that the experimental $0^+_2$ and $0^+_3$ states can 
not be described within $^{16}$O+$\alpha$
cluster models because they are not simple $^{16}$O+$\alpha$ cluster states.
The cluster-GCM calculation shows reasonable results for the energy 
levels except for the $6^+$-$8^+$ level spacing as already shown in preceding works 
with $^{16}$O+$\alpha$ cluster models \cite{Matsuse75,Nemoto72}.
The AMD(VAP) and hybrid calculations reproduce the ground band spectra.
In particular, 
the small level spacing between $6^+$ and $8^+$ states 
is described well by the cluster breaking component 
in the $8^+$ state at the band terminal consistently with the results of 
the cranking AMD calculation \cite{KanadaEnyo:1994kw}.

\begin{figure}[tb]
	\begin{center}
	\includegraphics[width=14cm]{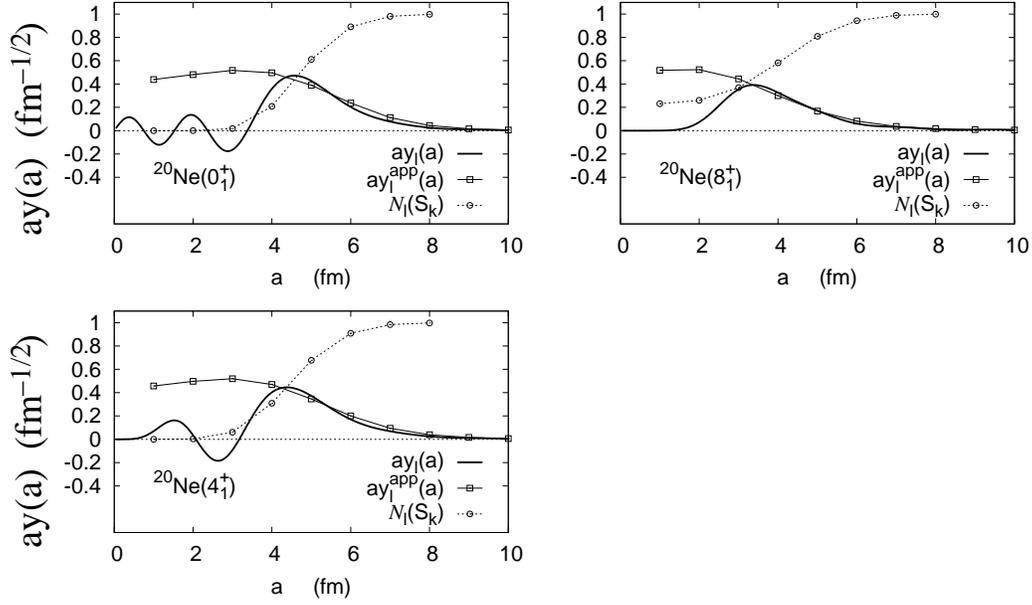}
    \end{center}
  \caption{\label{fig:20Ne-gcm}
The approximated RWA $ay^{\rm app}_l(a)$ and the exact $ry_l(a)$
of the $^{16}$O+$\alpha$ channel for the $J^\pi=0^+$, $4^+$, and $8^+$ states 
in the ground band 
of $^{20}$Ne 
obtained by 
the cluster-GCM calculation.
The allowedness factor $\mathcal{N}_{l}(S_k=a)$, which indicates
the weakness of the antisymmetrization effect of nucleons between clusters, in 
the projected BB wave function $\Phi^{J\pi}_{\rm BB}(S_k)$ is also shown.}
\end{figure}

\begin{figure}[tb]
	\begin{center}
	\includegraphics[width=14cm]{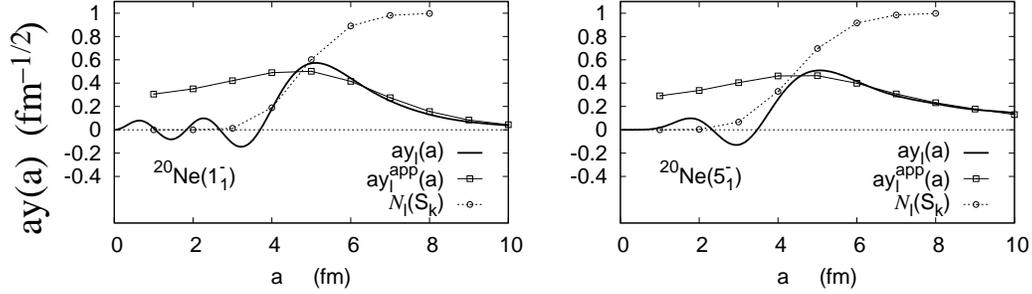}
    \end{center}
    \caption{\label{fig:20Ne-gcm-np}
Same as Fig.~\ref{fig:20Ne-gcm}
but for the $J^\pi=1^-$ and $5^-$ states in the $K^\pi=0^-$ band of $^{20}$Ne.}
\end{figure}

\begin{figure}[tb]
	\begin{center}
	\includegraphics[width=14cm]{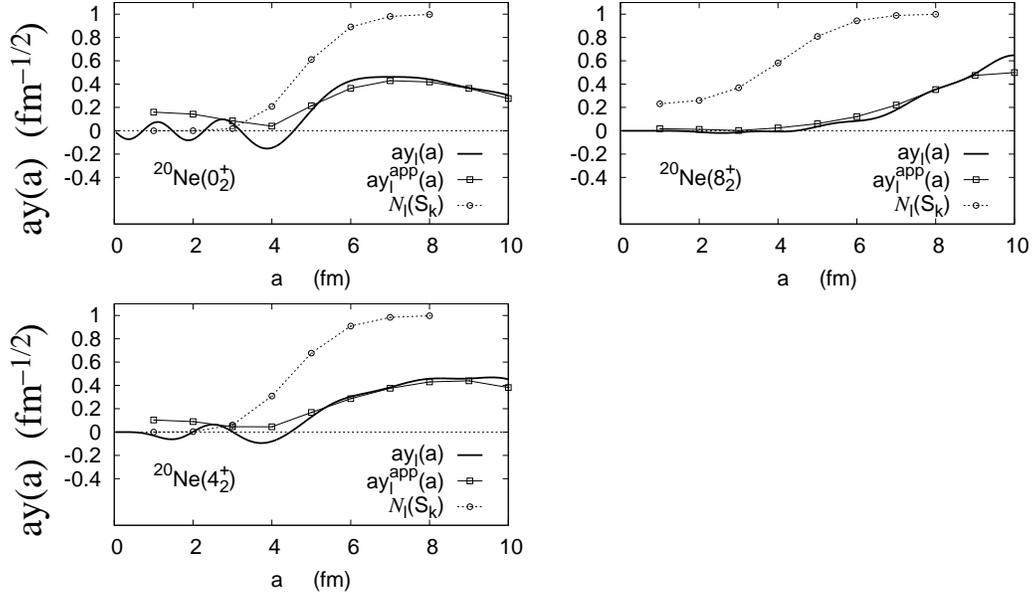}
    \end{center}
  \caption{\label{fig:20Ne-gcm2}
Same as Fig.~\ref{fig:20Ne-gcm}
but for the $J^\pi=0^+$, $4^+$, and $8^+$ states in the higher-nodal 
 $K^\pi=0^+$ band of $^{20}$Ne.}
\end{figure}

We first discuss the results of the cluster-GCM calculation.
In Figs.~\ref{fig:20Ne-gcm}, \ref{fig:20Ne-gcm-np}, and 
\ref{fig:20Ne-gcm2}, 
the approximated RWA $ay_l^{\rm app}(a)$ for $\Phi^{\rm GCM}$ are compared 
with the exact values of $ay_l(a)$. 
The approximated RWA reasonably agrees with $ay_l(a)$ for bound states and resonance states in the region outer than the surface peak. 
The $6^+_2$ and $8^+_2$ states obtained 
by the cluster-GCM calculation have a feature of non-resonant 
continuum states, for which the approximation also works 
in the outer region. 

In the small $r$ region, the method of the approximated RWA does not work
because the antisymmetrization effect of nucleons between clusters
is rather strong and the norm overlap with a BB wave function does not directly indicate 
the $\alpha$ cluster probability at the certain position. 
We can judge the strength of the antisymmetrization effect using 
the allowedness factor ${\cal N}_l(S_k)$ shown in 
Figs.~\ref{fig:20Ne-gcm}, \ref{fig:20Ne-gcm-np}, and 
\ref{fig:20Ne-gcm2}. 
In the present result, it is found that 
the approximated RWA is not reliable for a small channel radius 
$S_k=a$ with ${\cal N}_l(S_k) < 0.4$ because of the strong antisymmetrization effect.
To reject the unreliable region with the strong antisymmetization effect,
we put a more severe condition ${\cal N}_l(S_k)\ge 0.6$ 
as the applicable region because the agreement of $ry_l^{\rm app}(r)$ to $ry_l(r)$ 
is rather well in the outside of the surface peak. 
Moreover, when the RWA is much smaller than the peak amplitude, 
the error becomes large even in the long distance tail part.
Therefore, we reject $ry_l^{\rm app}(r)$ if it is 
less than a half of the maximum amplitude in the  applicable region.
When the RWA has a broad peak in the  applicable region, 
the channel radius $a$ should be chosen around the peak position
as shown in the result for higher-nodal band memers,  $0^+_2$, $2^+_2$, 
and $4^+_2$ states.

\begin{table}[ht]
\caption{Ratios $ay_l^{\rm app}(a)/ry_l(a)$ 
of the approximated RWA to the exact RWA of 
the cluster-GCM calculation of $^{20}$Ne. The 
channel radii $a=5$ and 6 fm are chosen for the $K^\pi=0^+_1$ and $K^\pi=0^-$ bands
except for the $J^\pi=8^+_1$ state, and $a=6$ and 7 fm for the higher-nodal 
$K^\pi=0^+$ band.)
\label{tab:20Ne-ratio}}
\begin{center}
\begin{tabular}{|c|cccc|}
\hline
	&	$a=4$ 	&	$a=5$ 	&	$a=6$ 	&	$a=7$	 \\
	\hline
$^{20}$Ne($0^+_1$)	&		&	0.91 	&	1.16 	&		\\
$^{20}$Ne($2^+_1$)	&		&	0.92 	&	1.17 	&		\\
$^{20}$Ne($4^+_1$)	&		&	0.93 	&	1.18 	&		\\
$^{20}$Ne($6^+_1$)	&		&	0.95 	&	1.21 	&		\\
$^{20}$Ne($8^+_1$)	&	0.92 	&	0.98 	&		&		\\
$^{20}$Ne($1^-_1$)	&		&	0.88 	&	0.94 	&		\\
$^{20}$Ne($3^-_1$)	&		&	0.89 	&	0.94 	&		\\
$^{20}$Ne($5^-_1$)	&		&	0.91 	&	0.96 	&		\\
$^{20}$Ne($7^-_1$)	&		&	0.89 	&	0.91 	&		\\
$^{20}$Ne($0^+_2$)	&		&	&	0.85 	&	0.93 	\\
$^{20}$Ne($2^+_2$)	&		&	&	0.87 	&	0.94 	\\
$^{20}$Ne($4^+_2$)	&		&	&	0.94 	&	0.98 	\\
\hline
\end{tabular}
\end{center}
\end{table}

The ratios $y^{\rm app}_l(r)/y_l(r)$ of the approximated RWA 
to the exact RWA are listed in Table \ref{tab:20Ne-ratio}.
For the $K^\pi=0^+_1$ and $K^\pi=0^-$ 
band members except for the $J^\pi=8^+_1$ state,
the channel radii $a=5$ and 6 fm are chosen  because 
the amplitude in the applicable region satisfying the condition 
${\cal N}_l(S_k)\ge 0.6$ is maximum at $a=5$ fm.
For the $J^\pi=8^+_1$ state, the applicable region is $a\ge 4$ fm and 
the amplitude at $a=6$ fm is much smaller than the maximum amplitude at $a=4$ fm,
and therefore we choose the channel radii $a=4$ and 5 fm.
For the higher-nodal $K^\pi=0^+$ band, we choose larger channel radii $a=6$ and 7 fm 
as the peak position of the RWA shifts to the outer region around $a=7$-8 fm.
With the criterion that the allowedness factor should be ${\cal N}_l(S_k) \ge 0.6$
and the channel radius near the peak position should be chosen, 
we get good approximation of the approximated RWA with the exact value
within about 20\% error.

\begin{figure}[tb]
	\begin{center}
	\includegraphics[width=14cm]{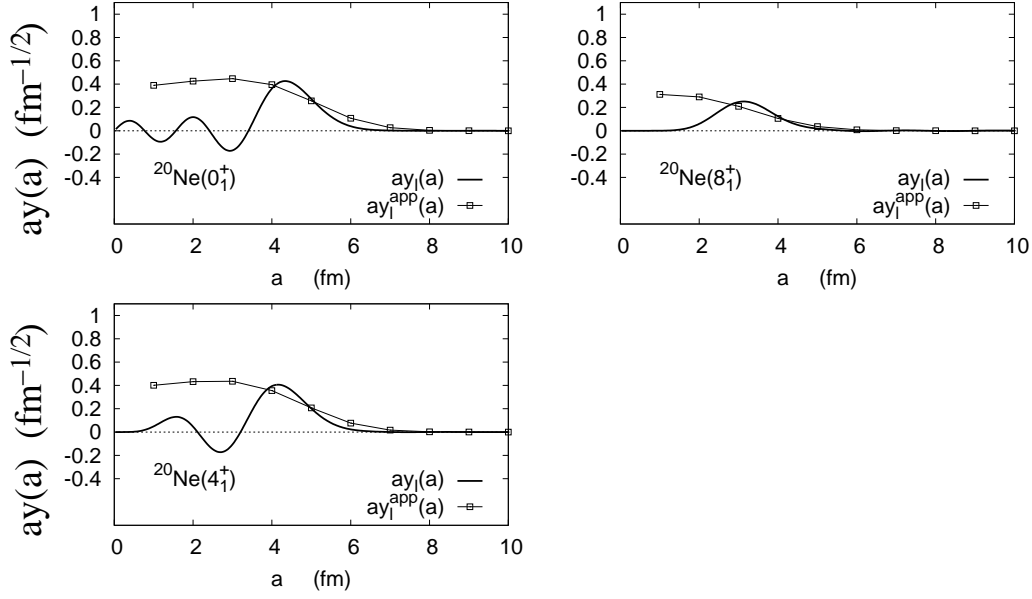}
    \end{center}
  \caption{\label{fig:20Ne-VAP-dia1}
The approximated RWA $ay^{\rm app}_l(a)$ and the exact RWA $ry_l(a)$
of the $^{16}$O+$\alpha$ channel for the $J^\pi=0^+$, $4^+$, and $8^+$ states in
the ground band of $^{20}$Ne obtained by the AMD(VAP) calculation.
}
\end{figure}

\begin{figure}[tb]
	\begin{center}
	\includegraphics[width=14cm]{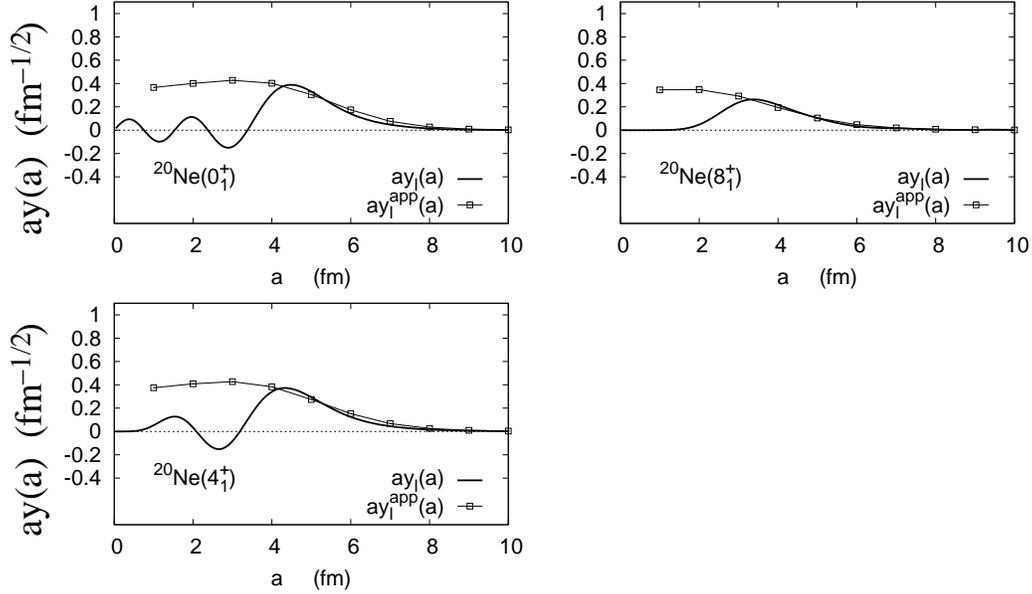}
    \end{center}
  \caption{\label{fig:20Ne-VAP-dia15}
Same as Fig.~\ref{fig:20Ne-VAP-dia1} but for 
the hybrid calculation of the AMD(VAP)+cluster wave functions.}
\end{figure}

We also check the approximation of the RWA for the AMD(VAP) wave functions 
and the hybrid AMD(VAP)+cluster wave functions. 
In the result for the AMD(VAP) wave functions shown in Fig.~\ref{fig:20Ne-VAP-dia1}, 
the approximation is not as good as the case of the cluster-GCM wave functions. 
As mentioned before, the AMD(VAP) wave function is the spin-parity eigen function 
projected from a single AMD wave function, and its inter-cluster 
wave function has a rapidly damping tail inconsistently 
with the correct asymptotic behavior. 
For such the localized function, the approximation 
does not work so well.
Instead, $ry^{\rm app}_l(r)$ corresponds to a smeared 
function of the exact $ry_l(r)$.
However, in the realistic situation, 
the inter-cluster wave function has an outer tail 
with the correct asymptotic behavior 
determined by the $\alpha$-decay energy, 
it should be a gradually changing function
for states near the threshold energy.
To describe the detailed behavior of the outer tail, 
we perform the hybrid calculation by 
superposing the AMD(VAP) wave functions for the $J^\pi=0^+,2^+,4^+,6^+$, and $8^+$ states 
and $^{16}$O+$\alpha$ cluster BB wave functions. 
In the hybrid wave functions, the tail parts of the inter-cluster wave functions are 
improved and it is found that 
the RWA $ay_l(a)$ can be approximated by $ay^{\rm app}(a)$ 
in the outer region
as shown in Fig.~\ref{fig:20Ne-VAP-dia15}.

It should be noted that the cluster component ${\cal P}^{\rm cluster}$
in the hybrid wave functions is less than 1 because of the cluster 
breaking component in the AMD(VAP) wave functions. 
The reduction effect of the cluster component to the RWA 
is properly taken into account 
in the present approximation of the RWA through the norm overlap.
The reduction is significant in the band terminal state $^{20}$Ne($8^+_1$) 
with ${\cal P}^{\rm cluster}\sim 0.44$ in the hybrid calculation.

\subsection{$\alpha$ decay widths of $^{20}$Ne} 

Using the relation (\ref{eq:rw-app}) based on the 
$R$-matrix theory of nuclear reaction, 
we can evaluate the dimensionless reduced width 
$\theta^2_\alpha(a)$ for the $\alpha$ decay with
the calculated RWA, $ay_l(a)$ and $ay^{\rm app}_l(a)$.

The theoretical values of $\theta^2_\alpha(a)$ calculated with the 
approximated RWA $ay^{\rm app}_l(a)$ are
shown in table \ref{tab:20Ne-drw} compared with those obtained with the 
exact RWA $ay_l(a)$. 
We choose the channel radius $a=5$ and 6 fm for 
the ground band and $K^\pi=0^-$ band members, and $a=6$ and 7 fm
for the higher-nodal states. 
In both cases of the cluster-GCM and the hybrid wave functions,
$\theta^2_\alpha(a)$ from $ay^{\rm app}_l(a)$ agrees
with that from $ay_l(a)$ within $20-30\%$ error. 
It means that the present approximation for the RWA is practically 
useful to evaluate the correct RWA at the channel radius in the region of 
our interest to estimate the $\alpha$-decay width.

\begin{table}[ht]
\caption{Dimensionless reduced width $\theta^2(a)$ of the 
$^{16}$O+$\alpha$ channel calculated with the relation 
\ref{eq:rw-app} using the 
exact RWA $ay_l(a)$ and the approximated one $ay^{\rm app}_l(a)$ for 
$^{20}$Ne obtained by the cluster-GCM calculation. The channel radii
$a=5$ and $a=6$ fm are chosen for the ground band and 
$K^\pi=0^-$ band members, and $a=6$ and $a=7$ fm are chosen for the higher-nodal states.
The energy $E$ (MeV) measured from the threshold is also listed.
\label{tab:20Ne-drw}}
\begin{center}
\begin{tabular}{|r|r|rr|rr|}
\hline
 \multicolumn{6}{|c|}{cluster-GCM} \\
&	$E$	&	 \multicolumn{2}{c|}{$\theta^2(a)$} &	 \multicolumn{2}{c|}{$\theta^2(a)$}\\
&  & \multicolumn{2}{c|}{ exact} & \multicolumn{2}{c|}{approx.}  \\
&	&	$a=5$ 	&		$a=6$ & 	$a=5$ 	&	$a=6$  \\
\hline
$^{20}$Ne($0^+_1$)	&	$-4.57$ 	&	0.30 	&	0.08 	&	0.25 	&	0.11 	\\
$^{20}$Ne($2^+_1$)	&	$-3.58$ 	&	0.28 	&	0.08 	&	0.24 	&	0.10 	\\
$^{20}$Ne($4^+_1$)	&	$-1.30$ 	&	0.23 	&	0.06 	&	0.20 	&	0.08 	\\
$^{20}$Ne($6^+_1$)	&	2.13 	&	0.15 	&	0.03 	&	0.13 	&	0.05 	\\
$^{20}$Ne($8^+_1$)	&	6.31 	&	0.05 	&	0.007 	&	0.05 	&		\\
$^{20}$Ne($1^-_1$)	&	0.11 	&	0.54 	&	0.39 	&	0.42 	&	0.35 	\\
$^{20}$Ne($3^-_1$)	&	1.80 	&	0.52 	&	0.38 	&	0.41 	&	0.34 	\\
$^{20}$Ne($5^-_1$)	&	4.86 	&	0.43 	&	0.35 	&	0.36 	&	0.32 	\\
$^{20}$Ne($7^-_1$)	&	8.18 	&	0.39 	&	0.26 	&	0.31 	&	0.22 	\\
	&		&	$a=6$ 	&	$a=7$  	&	$a=6$  	&	$a=7$  	\\
$^{20}$Ne($0^+_2$)	&	3.01 	&	0.37 	&	0.50 	&	0.26 	&	0.43 	\\
$^{20}$Ne($2^+_2$)	&	3.68 	&	0.31 	&	0.46 	&	0.24 	&	0.41 	\\
$^{20}$Ne($4^+_2$)	&	5.14 	&	0.18 	&	0.34 	&	0.16 	&	0.33 	\\
\hline
 \multicolumn{6}{|c|}{hybrid AMD(VAP)+cluster} \\
 &	$E$	&	 \multicolumn{2}{c|}{$\theta^2(a)$} &	 \multicolumn{2}{c|}{$\theta^2(a)$}\\
&  & \multicolumn{2}{c|}{ exact} & \multicolumn{2}{c|}{approx.}  \\
	&	&	$a=5$ 	&		$a=6$  & 	$a=5$ 	&	$a=6$  \\
	\hline
$^{20}$Ne($0^+_1$)	&	$-6.08$ 	&	0.18 	&	0.040 	&	0.15 	&	0.061 	\\
$^{20}$Ne($2^+_1$)	&	$-4.79$ 	&	0.18 	&	0.039 	&	0.15 	&	0.059 	\\
$^{20}$Ne($4^+_1$)	&	$-2.24$ 	&	0.14 	&	0.031 	&	0.13 	&	0.047 	\\
$^{20}$Ne($6^+_1$)	&	1.79 	&	0.094 	&	0.020 	&	0.085 	&	0.029 	\\
$^{20}$Ne($8^+_1$)	&	5.22 	&	0.018 	&	0.002 	&	0.018 	&		\\
\hline
\end{tabular}
\end{center}
\end{table}

\begin{table}[ht]
\caption{The experimental and theoretical values of 
the dimensionless reduced width $\theta^2(a)$ of the 
$^{16}$O+$\alpha$ channel of $^{20}$Ne at the channel radius $a=5$ and $a=6$ fm.
The experimental $\theta^2(a)$ is calculated using 
the measured $\alpha$ decay widths \cite{Tilley98}. 
The theoretical values are those of the RGM and GCM calculations taken 
from Refs.~\cite{Matsuse75,Nemoto72}.
The energy $E$ (MeV) measured from the threshold is also listed.
\label{tab:20Ne-exp-drw}}
\begin{center}
\begin{tabular}{|r|rrr|rr|}
\hline
&  \multicolumn{5}{c|}{Exp.} \\
& 	$E$	 &  
\multicolumn{2}{c|}{$\theta^2(a)$ } &&$\Gamma_\alpha$ (keV)			\\
	&	&	$a=5$ 	&		$a=6$  & &	\\
	\hline
$^{20}$Ne($0^+_1$)	&	$-$4.73	&		&		&		&		\\
$^{20}$Ne($2^+_1$)	&	$-$3.1	&		&		&		&		\\
$^{20}$Ne($4^+_1$)	&	$-$0.48	&		&		&		&		\\
$^{20}$Ne($6^+_1$)	&	4.05	&	0.073(17)	&	0.0103(23)	&&	0.110(25)			\\
$^{20}$Ne($8^+_1$)	&	7.22	&	0.0095(27)	&	0.00094(27)	&&	0.035(10)\\
$^{20}$Ne($1^-_1$)	&	1.06	&	1.04 	&	0.32 	&&	0.028			\\
$^{20}$Ne($3^-_1$)	&	2.43	&	0.97 	&	0.28 	&&	8.2			\\
$^{20}$Ne($5^-_1$)	&	5.53	&	1.08 	&	0.32 	&&	145			\\
$^{20}$Ne($7^-_1$)	&	10.64	&	0.24 	&	0.07 	&&	110			\\
	&		&	$a=6$ 	&	$a=7$ 	&		&		\\
$^{20}$Ne($0^+_{\rm hn}$)	&	4	&	$>$0.39	&	$>$0.37	&&	$>$800			\\
$^{20}$Ne($2^+_{\rm hn}$)	&	4.3	&	$\sim$0.52	&	$\sim$0.43	&&	$\sim$800			\\
$^{20}$Ne($4^+_{\rm hn}$)	&	6.06	&	0.23 	&	0.17 	&&	350			\\
\hline
& \multicolumn{3}{c|}{RGM \cite{Matsuse75}} & \multicolumn{2}{c|}{GCM \cite{Nemoto72}} \\
	&		$E$	&	\multicolumn{2}{c|}{$\theta^2(a)$ }		&	$E$	&	$\theta^2(a)$	\\
	&	&	$a=5$ 	&	$a=6$ 	& &	$a=6$ 	\\
	\hline
$^{20}$Ne($0^+_1$)	&	$-$4.26	&		&		&	$-3.9$	&	0.057	\\
$^{20}$Ne($2^+_1$)	&	$-$3.25	&		&		&	$-2.72$	&	0.052	\\
$^{20}$Ne($4^+_1$)	&	$-$0.94	&		&		&	0.05	&	0.041	\\
$^{20}$Ne($6^+_1$)	&	2.52	&	0.49	&	0.054	&	3.73	&	0.024	\\
$^{20}$Ne($8^+_1$)	&	6.77	&	0.16	&	0.015	&	9.86	&	0.006	\\
$^{20}$Ne($1^-_1$)	&	0.3	&	2.2	&	0.57	&	$-0.3$	&	0.267	\\
$^{20}$Ne($3^-_1$)	&	1.98	&	2.23	&	0.58	&	1.69	&	0.265	\\
$^{20}$Ne($5^-_1$)	&	5.08	&	2.28	&	0.6	&	5.3	&	0.271	\\
$^{20}$Ne($7^-_1$)	&	9.89	&	2.28	&	0.61	&	9.91	&	0.298	\\
$^{20}$Ne($0^+_{\rm hn}$)	&		&		&		&	3.01	&	0.604	\\
$^{20}$Ne($2^+_{\rm hn}$)	&		&		&		&	3.77	&	0.578	\\
$^{20}$Ne($4^+_{\rm hn}$)	&		&		&		&	5.46	&	0.501	\\
\hline
\end{tabular}
\end{center}
\end{table}

In table \ref{tab:20Ne-exp-drw}, we list the experimental $\theta^2_\alpha(a)$ for 
resonance states obtained with the observed decay width 
$\Gamma_\alpha$. We also show the theoretical $\theta^2_\alpha(a)$
of the RGM calculation in Ref.~\cite{Matsuse75} and the GCM calculation in Ref.~\cite{Nemoto72}.
The GCM calculation in Ref.~\cite{Nemoto72} is a bound state approximation and 
the relation (\ref{eq:rw-app}) of the $\alpha$-decay width and the RWA 
is used. The calculation is quite similar to the 
present calculation but the interaction used in Ref.~\cite{Nemoto72} is 
different from the present one.
In the RGM calculation in Ref.~\cite{Matsuse75}, the 
$\theta^2_\alpha(a)$ is evaluated by the
phase shift analysis by solving the scattering problem.

The present result of $\theta^2_\alpha(a)$ obtained by the cluster-GCM calculation is
similar to those of Refs.~\cite{Matsuse75,Nemoto72}.
The theoretical $\theta^2_\alpha(a)$ is comparable 
to the experimental data. 
There are significant disagreements between calculated values and experimental ones 
for the decay width of 
$^{20}$Ne($8^+_1$) and $^{20}$Ne($7^-$).  
For those states, the cluster-GCM calculation overestimates 
the experimental $\alpha$-decay width by a factor $2-5$ 
as well as the cluster-model calculations in Refs.~\cite{Matsuse75,Nemoto72}.
The result is improved in the hybrid calculation 
where the cluster component ${\cal P}^{\rm cluster}$ of $^{20}$Ne($8^+_1$) 
reduces to ${\cal P}^{\rm cluster}\sim 0.4$ because of the mixing of the cluster breaking component.

According to the $R$-matrix theory, the relation (\ref{eq:rw-app}) of the reduced width and the RWA is a good approximation, especially, for narrow resonances. However, 
strictly speaking, it is not necessarily good for broad resonances. 
Nevertheless, the present result using 
(\ref{eq:rw-app}) in the bound state approximation
shows reasonable values of the $\alpha$-decay width 
even for such broad resonances as $^{20}$Ne($0^+_{\rm hn}$) and 
 $^{20}$Ne($2^+_{\rm hn}$). It may suggest that the bound state approximation is still useful 
for a rough estimation of cluster-decay width. 

In the present result of $^{20}$Ne it is found that the $ay^{\rm app}_l(a)$ 
is a good approximation of the RWA at the surface region and it is useful 
for our aim to give qualitative discussion of the $\alpha$-decay width. 
 
\subsection{$^8$Be}
\begin{figure}[tb]
	\begin{center}
	\includegraphics[width=14cm]{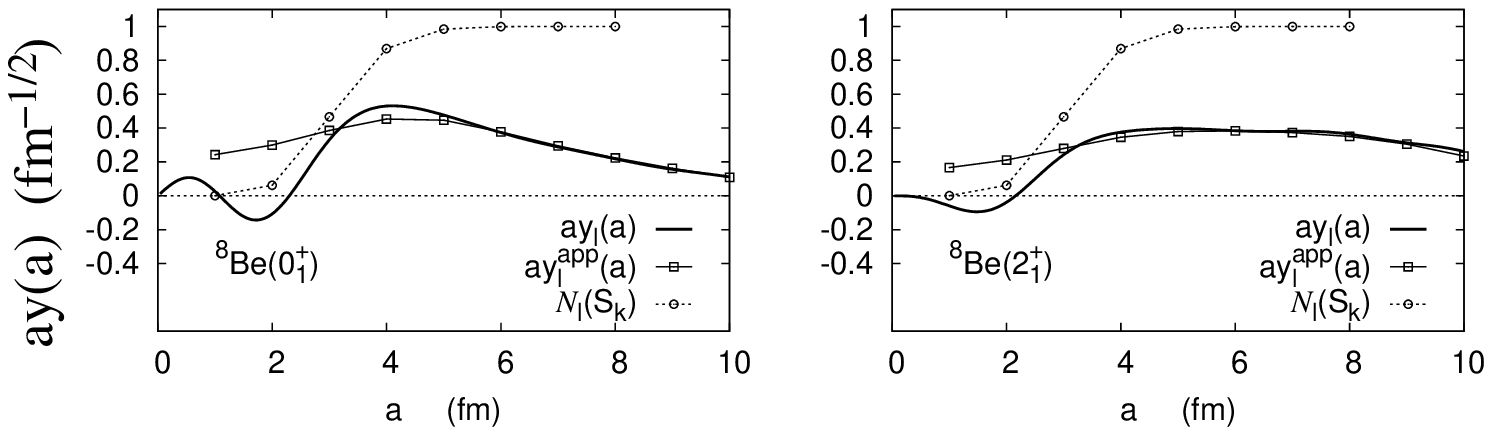}
    \end{center}
  \caption{\label{fig:8Be-gcm}
The approximated RWA $ay^{\rm app}(a)$ and the exact $ry_l(a)$
of the $\alpha$+$\alpha$ channel in the $0^+_1$ and $2^+_1$ states of 
$^{8}$Be obtained by 
the cluster-GCM calculation using the Volkov No.2 force ($m=0.60$) and the
width parameter $\nu=0.25~\mathrm{fm}^{-2}$. 
The allowedness factor $\mathcal{N}_{l}(S_k=a)$, which indicates
the weakness of the antisymmetrization effect between clusters, in 
the projected BB wave function $\Phi^{J\pi}_{\rm BB}(S_k)$ with $S_k=a$ is also shown.}
\end{figure}

We perform the similar analysis 
for $^8$Be($0^+_1$) and  $^8$Be($2^+_1$) obtained 
by the $\alpha$+$\alpha$ cluster-GCM calculation and 
check the applicability of the approximated RWA. 
The Volkov No.2 interaction with $m=0.60$ is used to reproduce the energy $E$
of $^8$Be($0^+_1$). For the basis wave functions in the cluster-GCM calculation, 
the $\alpha$+$\alpha$ BB wave functions
with $S_k=1, 2, \cdots, 10$ fm are used, and the width parameter $\nu=0.25$ fm$^{-2}$ is chosen for the 
$\alpha$ cluster wave function.
The approximated RWA $ry^{\rm app}_l(r)$ is shown in  Fig.~\ref{fig:8Be-gcm} compared with the correct RWA.
It is shown that $ry^{\rm app}_l(r)$ is a good approximation to describe the RWA 
of the tail part because $^8$Be($0^+_1$) and $^8$Be($2^+_1$) are quasi-bound $\alpha$+$\alpha$ 
states having the long tail of the inter-cluster wave function.
In table \ref{tab:8Be-width}, we show 
the dimensionless reduced $\alpha$-decay width $\theta^2(a)$ of $^8$Be 
calculated with the relation (\ref{eq:rw-app}) using the exact RWA ($ay_l(a)$) and the approximated one ($ay^{\rm app}_l(a)$). 
The agreement of $\theta^2(a)$ evaluated with $ay^{\rm app}_l(a)$
with that using $ay_l(a)$ is rather good with 20\% error at most.
Compared with the experimental $\theta^2(a)$ given by the measured $\alpha$-decay width $\Gamma_\alpha$, 
it is found that the calculation reasonably describes the experimental decay width of $^8$Be($0^+_1$). 
Even for the case of the broad resonance of $^8$Be($2^+_1$), 
the $\alpha$-decay width is reasonably described by the calculation.

\begin{table}[ht]
\caption{Calculated energies $E$ (MeV) measured from the $2\alpha$ threshold 
and dimensionless reduced  width $\theta^2(a)$ of the $\alpha$+$\alpha$ channel 
for $^{8}$Be($0^+_1$) and $^{8}$Be($2^+_1$) obtained by the cluster-GCM calculation
compared with the experimental data \cite{Kelley04}.
The calculated $\theta^2(a)$ is evaluated with the relation \ref{eq:rw-app}
using the exact 
RWA $ay_l(a)$ and the approximated one $ay^{\rm app}_l(a)$.
The channel radius $a=5$, 6, and 7 fm are chosen.
\label{tab:8Be-width}}
\begin{center}
\begin{tabular}{|c|c|ccc|ccc|}
\hline
&   \multicolumn{7}{c|}{cluster-GCM} \\
&$E$  &  \multicolumn{3}{c|}{$\theta^2(a)$} &  \multicolumn{3}{c|}{$\theta^2(a)$}\\
&  &  \multicolumn{3}{c|}{exact} & \multicolumn{3}{c|}{approx.}  \\
	&	&	$a=5$ 	&	$a=6$ 	&	$a=7$ 	&	$a=5$	&	$a=6$ &	$a=7$ 	\\
	\hline
$^{8}$Be($0^+_1$)	&	0.18	&	0.33 	&	0.28 	&	0.20 	&	0.38 	&	0.28 	&	0.19 	\\
$^{8}$Be($2^+_1$)	&	2.37	&	0.24 	&	0.29 	&	0.33 	&	0.26 	&	0.30 	&	0.34 	\\
\hline
&   \multicolumn{7}{c|}{Exp.} \\
&$E$  &  \multicolumn{3}{c|}{$\theta^2(a)$}&  $\Gamma_\alpha$(keV)	& & \\ 
	&	&	$a=5$ 	&	$a=6$ 	&	$a=7$ 	& & & \\
	\hline
$^{8}$Be($0^+_1$)	&	0.092	&	0.27 	&	0.19 	&	0.14 	&	0.00557	&		&		\\
$^{8}$Be($2^+_1$)	&	3.122	&	0.50 	&	0.47 	&	0.48 	&	1513	&		&		\\
\hline	
\end{tabular}
\end{center}
\end{table}

\subsection{$^{9}$Li}
We apply the present approximation to $^9$Li and estimate the 
$t$-decay width of the $^6$He+$t$ cluster resonances predicted 
in the previous work \cite{KanadaEn'yo:2011nc}.
The present approximation is applicable to the 
cluster channel $^6$He($0^+$)+$t$ where the orbital angular momentum 
of the inter-cluster motion is decoupled from the internal spins of clusters.    
$^9$Li wave functions are obtained by 
the $^6$He+$t$ cluster-GCM calculation in the same way 
as Ref.~\cite{KanadaEn'yo:2011nc}.
Namely, the $^6$He+$t$-cluster BB wave functions with $S_k=1,\cdots,8$ fm
are superposed to describe $J^\pi=1/2^-$, $3/2^-$, $5/2^-$, and $7/2^-$ states of $^9$Li.
Practically, the cluster wave functions are described by the linear combination of 
AMD wave functions with specific configurations as done in the previous work.
The $^6$He cluster is expressed by the H.O. shell-model configurations. The configuration 
mixing in the major shell is taken into account, and all 
$^6$He($0^+$) and $^6$He($2^+$) states in the $(0s)^4(0p)^2$ configurations 
are incorporated. 
In the ground state, $^6$He($0^+_1$), obtained in the $p$-shell, 
$(p_{3/2})^2$ and $(p_{1/2})^2$ configurations are mixed.
Because of the configuration mixing in the $^6$He cluster,
it is not easy to get the RGM norm kernel and to calculate the exact RWA 
of the $^6$He($0^+_1$)+$t$ cluster channel.
Instead, we calculate the overlap norm of the $^9$Li wave function with the 
$^6$He($0^+_1$)+$t$-cluster BB wave function at a certain channel radius $S_k=a$ 
and obtain the approximated value $ay^{\rm app}_l(a)$ of the RWA to discuss 
the $t$-decay width of the cluster resonance states.

The interaction and width parameters are those used in the previous work. 
The interaction is Volkov No.2 with $m=0.60,b=h=0.125$ supplemented by 
the spin-orbit term of the G3RS force with the strength $u_I=-u_{II}=1600$ MeV, 
which is adjusted to reproduce the energy spectra of $^{10}$Be
with the $^6$He+$\alpha$ cluster-GCM calculation. 
The width parameter $\nu=0.235$ fm$^{-2}$ is used.
In the present work, all $K$ states are mixed in the cluster-GCM calculation while 
$K$ was truncated as $|K|\le 3/2$ in the previous calculation. 
For $J^\pi=5/2^-$ and $7/2^-$ states, the lower energy spectra is changed by 
the mixing of high $K$ states, but it does not change 
the feature of the $^6$He+$t$ resonance states near the threshold.

\begin{figure}[tb]
\begin{center}
	\includegraphics[width=6.5cm]{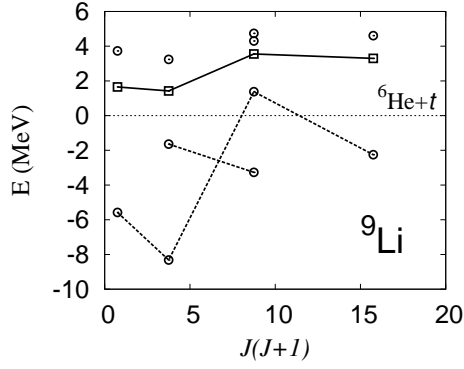} 
\end{center}
\vspace{1cm}	
  \caption{\label{fig;li9-ene}
Calculated energy spectra of negative-parity states in $^9$Li 
obtained by the $^{6}$He+$t$ cluster-GCM calculation
using the Volkov No.2 force ($m=0.60,b=h=0.125$) supplemented by 
the spin-orbit term of the G3RS force ($u_I=-u_{II}=1600$ MeV).
The energies $E$ of $^9$Li states 
measured from the $^{6}$He+$t$ threshold are plotted as functions of 
the spin $J(J+1)$.}
\end{figure}

We here briefly explain the structure of the ground and excited states of 
$^9$Li obtained by the  $^6$He+$t$ cluster-GCM calculation.
For more details of the structure of $^9$Li, the reader is referred to 
Ref.~\cite{KanadaEn'yo:2011nc}.
The energies $E$ measured from the $^6$He+$t$ threshold energy 
are plotted as function of the spin $J(J+1)$ in Fig.~\ref{fig;li9-ene}. 
$^9$Li($1/2^-_1$), $^9$Li($3/2^-_1$), $^9$Li($5/2^-_2$),
$^9$Li($7/2^-_1$) are regarded as members of the ground band.
In the highly excited states near the threshold, 
$^9$Li($1/2^-_2$), $^9$Li($3/2^-_3$), $^9$Li($5/2^-_3$),
$^9$Li($7/2^-_2$) show the $^{6}$He and $t$ resonance feature and 
they are regarded as the $^{6}$He+$t$ cluster resonances 
as discussed in the previous work. 
$^9$Li($3/2^-_2$) and $^9$Li($5/2^-_1$) are
shell model-like states given by $p$-shell configurations having dominantly 
the $K=3/2$ ($L_z=2$) component.
$^9$Li($1/2^-_3$), $^9$Li($3/2^-_4$), $^9$Li($5/2^-_5$), and
$^9$Li($7/2^-_3$) are considered to be non-resonance states 
strongly coupling with $^{6}$He+$t$ continuum states. 

Using the overlap norm of the $^9$Li wave function with the 
$^6$He($0^+_1$)+$t$-cluster BB wave function at a certain channel radius $S_k=a$, 
we calculate the approximated value $ay^{\rm app}_l(a)$ of the RWA and estimate 
the partial decay width for the $^6$He($0^+_1$)+$t$ channel of the resonances near the threshold.
In the present $^6$He+$t$ cluster-GCM calculation, 
the channel coupling is incorporated, and therefore, 
$^6$He($0^+$)+$t$ and $^6$He($2^+$)+$t$ cluster channels are coupled in
$^9$Li wave functions. However, the present approximation of the RWA is applicable 
only for the case 
that the relative angular momentum $l$ does not couple with 
the intrinsic angular-momenta of clusters. Therefore, we can analyze only the 
$^6$He($0^+_1$)+$t$-cluster component and discuss the RWA and the partial 
decay width of this channel.
As mentioned before, 
the cluster wave function for the ground state $^6$He($0^+_1$) 
is the linear combination of H.O.
$(p_{3/2})^2$ and $(p_{1/2})^2$ coupling to totally zero angular momentum.
In the $^6$He($0^+_1$)+$t$-cluster BB wave function with the distance parameter 
$S_k$, which is expressed by the linear combination of AMD wave functions, 
this corresponds to the sub-projection 
(spin-parity projection of the subsystem $^6$He) and the state mixing 
in the $^6$He cluster.

\begin{figure}[tb]
\begin{center}
	\includegraphics[width=6.5cm]{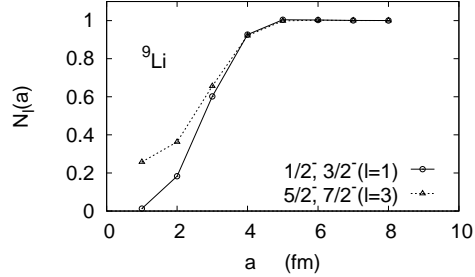}
\end{center} 
\vspace{1.cm}
  \caption{\label{fig:li9-N} 
The allowedness factor $\mathcal{N}_{l}(S_k=a)$, which indicates
the weakness antisymmetrization effect between clusters, in 
the $^6$He+$t$-cluster wave function with $S_k=a$.}
\end{figure}

We first determine the applicable region of the present approximation of the RWA by excluding 
the channel radius $S_k=a$ with the strong antisymmetrization effect.
From the calculated allowedness factor 
${\cal N}_l(S_k)$ of $l=1$ for
$J^\pi=1/2^-$ and $J^\pi=3/2^-$ states and that of $l=3$ for 
$J^\pi=5/2^-$ and  $J^\pi=7/2^-$ states 
shown in Fig.~\ref{fig:li9-N}, 
we find that the region $S_k \ge 3$ fm satisfies
the criterion ${\cal N}_l(S_k)\ge 0.6$ and consider 
this region as the applicable region of the present approximation.

\begin{figure}[tb]
\begin{center}
	\includegraphics[width=14cm]{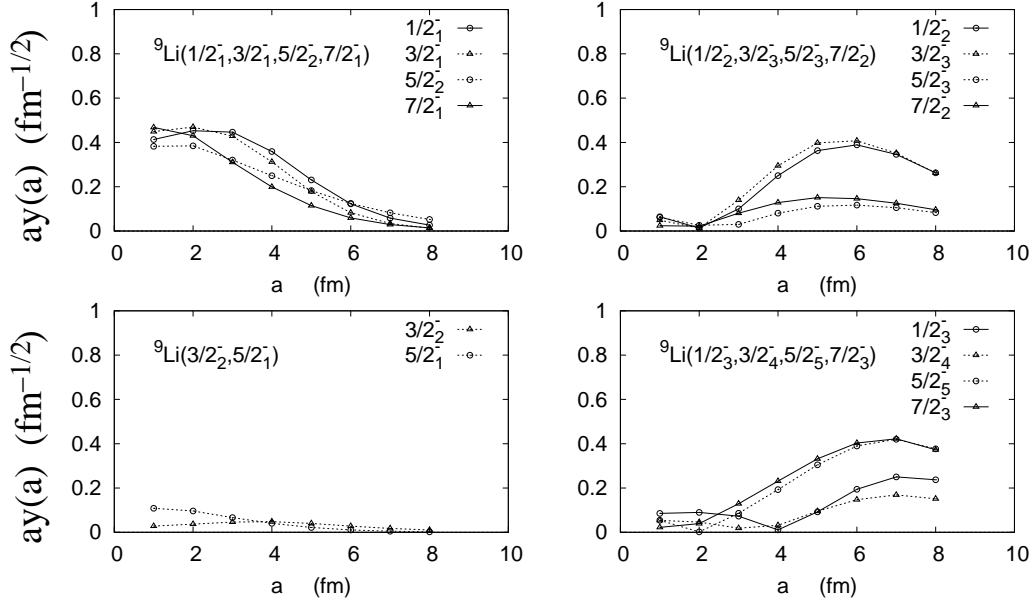}
\end{center} 
  \caption{\label{fig:9Li-rwa-app}
The approximated RWA
    $ay^{\rm app}_l(a)$ of the $^{6}$He($0^+_1$)-$t$ channel in 
$^9$Li obtained by the $^{6}$He+$t$ cluster-GCM calculation.
Upper left panel: $ay^{\rm app}_l(a)$ for the ground band members: 
$^9$Li($1/2^-_1$), $^9$Li($3/2^-_1$), $^9$Li($5/2^-_2)$,
and $^9$Li($7/2^-_1$). 
Lower left panel: that for the $K=3/2^-$ band members, $^9$Li($5/2^-_1)$ and
$^9$Li($3/2^-_2$).
Upper right panel: that for the $^{6}$He+$t$ resonance states,
$^9$Li($1/2^-_2$), $^9$Li($3/2^-_3$), $^9$Li($5/2^-_3)$, and
$^9$Li($7/2^-_2$).
Lower right panel: that 
$^9$Li($1/2^-_3$), $^9$Li($3/2^-_4$), $^9$Li($5/2^-_5)$,
$^9$Li($7/2^-_3$) which are regarded as 
non-resonant states strongly coupling with continuum states. 
}
\end{figure}

The calculated $ay^{\rm app}_l(a)$ for the $^6$He($0^+_1$)+$t$ channel in the cluster-GCM 
wave functions of the ground and excited states of $^9$Li is shown in Fig.~\ref{fig:9Li-rwa-app}.
In the ground band members, the amplitude at $a=3\sim 4$ fm 
indicates the relatively large probability of the $t$ cluster at the surface 
in $^9$Li($1/2^-_1$) and  $^9$Li($3/2^-_1$) compared with 
 $^9$Li($5/2^-_2$) and $^9$Li($7/2^-_1$). The surface probability of $t$ is 
suppressed in the high spin $7/2^-_1$ states, maybe, because of the centrifugal barrier.
For $^9$Li($5/2^-_2$), $ay^{\rm app}_l(a)$
shows a long tail of the inter-cluster wave function reflecting the 
energy position $E$ near the threshold.  
In $^9$Li($3/2^-_2$) and $^9$Li($5/2^-_1$), $ay^{\rm app}_l(a)$ is very small. 
This is consistent with the fact that these states are dominated by the 
$K^\pi=3/2^-$ states and mainly contain the excited cluster component $^6$He($2^+$)
rather than $^6$He($0^+$).
$^9$Li($1/2^-_2$), $^9$Li($3/2^-_3$), $^9$Li($5/2^-_3$), and 
$^9$Li($7/2^-_2$) show the peak structure of the RWA around $a=6$ fm
indicating the resonance feature of developed $^6$He and $t$ clusters.
The smaller RWA values in $^9$Li($5/2^-_3$) and 
$^9$Li($7/2^-_2$) than those in $^9$Li($1/2^-_2$) 
and $^9$Li($3/2^-_3$) are understood by the coupling with 
the $^6$He($2^+$)+$t$ with the $l=1$ wave of relative motion because of the 
alignment of the $^6$He cluster in high spin states.
In non-resonant continuum states, $^9$Li($1/2^-_3$), $^9$Li($3/2^-_4$), $^9$Li($5/2^-_5$),
$^9$Li($7/2^-_3$), 
$ay^{\rm app}_l(a)$ is no longer confined in the finite region.

We estimate the partial width of $^6$He($0^+_1$)+$t$ decay with the relation
(\ref{eq:rw-app}) using 
the calculated $ay^{\rm app}_l(a)$. Experimentally, 
the $^6$He+$t$ resonances have not been observed yet.
We here use the theoretical values of the decay energy $E$ in the estimation of the decay width.
The calculated dimensionless reduced width $\theta^2(a)$ and the decay width $\Gamma$ are listed 
in Table \ref{tab:9Li-width}. We choose the channel radius $a=3,4,$ and 5 fm for the ground band
and $a=5,6,$ and 7 fm for the cluster resonances.
The calculated partial decay width $\Gamma_{^6{\rm He}(0^+_1){\rm -}t}$ 
of $^9$Li($5/2^-_2$) is as small as 10 keV order 
because this state is the shell model state with less cluster development.
For the cluster resonances, $^9$Li($1/2^-_2$) and $^9$Li($3/2^-_3$), 
the present result suggests the width 
$\Gamma_{^6{\rm He}(0^+_1){\rm -}t}$ of the order 1 MeV, which is consistent with the 
width estimation of the pseudo potential method in the previous work.
Much smaller partial widths are suggested for $^9$Li($5/2^-_3$) and 
$^9$Li($7/2^-_2$) because $^6$He($0^+_1$)+$t$ component is suppressed 
originating in the coupling with the $l=1$-wave $^6$He($2^+$)+$t$ 
channel due to the $^6$He alignment.
In the present calculation, 
the $^6$He($2^+_1$)+$t$ channel is open for $^9$Li($5/2^-_3$) and 
$^9$Li($7/2^-_2$) while it is closed for $^9$Li($1/2^-_2$) 
and $^9$Li($3/2^-_3$). For the total $t$-decay width of $^9$Li($5/2^-_3$) and 
$^9$Li($7/2^-_2$), it is necessary to estimate also 
the $^6$He($2^+_1$)+$t$ decay width. However, since the application of the 
present approximation is restricted only for the spinless cluster case, 
it is a future problem to be solved.

In the present calculation, we assume the
H.O. $p$-shell configuration for the $^6$He cluster. 
Although such the H.O. $^6$He wave function is too simple to describe the details of 
the $^6$He structure, it may have a significant overlap with more sophisticated $^6$He 
wave function and therefore
the present calculation may be useful for order estimation.

It should be also noted that the $n$ decay is important to discuss the 
total width of $^9$Li states. The $n$ decay 
channel is omitted in the present $^6$He+$t$ cluster-GCM calculation. However, for
the $^6$He+$t$ cluster resonances, $^9$Li($1/2^-_2$), $^9$Li($3/2^-_3$), 
$^9$Li($5/2^-_3$), and $^9$Li($7/2^-_2$), the $n$ decay might be suppressed 
because the $^6$He+$t$ cluster structure develops so well that 
those cluster states have small overlap with 
the $^8$Li+$n$ component and hence it is naively expected that the 
$t$ decay can be the dominant decay channel.  Of course, it is not the case if 
the energy position of the $^6$He+$t$ cluster states is low enough to close the 
$t$ decay channel.

\begin{table}[ht]
\caption{The calculated partial decay width and dimensionless reduced width $\theta^2(a)$ 
for the $^{6}$He($0^+_1$)+$t$ decay of $^9$Li evaluated with $ay^{\rm app}_l(a)$. 
The channel radius $a=3$, 4, and 5 fm are chosen for the ground band members and 
 $a=5$, 6, and 7 fm are chosen for the $^{6}$He+$t$ resonance states near the threshold.
The energy $E$ (MeV) measured from the $^{6}$He+$t$ threshold is also listed. 
\label{tab:9Li-width}}
\begin{center}
\begin{tabular}{|c|c|ccc|ccc|}
\hline
& $E$  &  \multicolumn{3}{c|}{$\theta^2(a)$ } &
 \multicolumn{3}{c|}{$\Gamma_{^6{\rm He}(0^+_1){\rm -}t}$ (MeV) }  \\
&  &   \multicolumn{3}{c|}{approx.} & & & \\
	&		&	$a=3$	&	$a=4$ 	&	$a=5$ 	&	$a=3$ 	&	$a=4$ 	&	$a=5$ 	\\
	\hline
$^{9}$Li($1/2^-_1$)	&	$-$5.58 	&	0.20 	&	0.17 	&	0.089 	&		&		&		\\
$^{9}$Li($3/2^-_1$)	&	$-$8.32 	&	0.18 	&	0.13 	&	0.053 	&		&		&		\\
$^{9}$Li($5/2^-_2$)	&	1.38 	&	0.10 	&	0.083 	&	0.055 	&	0.002 	&	0.005 	&	0.009 	\\
$^{9}$Li($7/2^-_1$)	&	$-$2.25 	&	0.10 	&	0.053 	&	0.022 	&		&		&		\\
& & & & & & & \\													
	&	&	$a=5$ 	&	$a=6$ 	&	$a=7$ 	&	$a=5$ 	&	$a=6$ 	&	$a=7$ 	\\
	\hline
$^{9}$Li($1/2^-_2$)	&	1.65 	&	0.22 	&	0.30 	&	0.28 	&	0.68 	&	0.88 	&	0.75 	\\
$^{9}$Li($3/2^-_3$)	&	1.42 	&	0.26 	&	0.33 	&	0.29 	&	0.70 	&	0.84 	&	0.68 	\\
$^{9}$Li($5/2^-_3$)	&	3.55 	&	0.021 	&	0.027 	&	0.026 	&	0.045 	&	0.074 	&	0.077 	\\
$^{9}$Li($7/2^-_2$)	&	3.30 	&	0.038 	&	0.043 	&	0.037 	&	0.071 	&	0.10 	&	0.10 	\\
\hline
\end{tabular}
\end{center}
\end{table}

\section{Summary and outlooks}\label{sec:summary}
We proposed a 
method to approximately evaluate the RWA 
of the spinless two-body cluster channel 
using the overlap with the BB cluster wave function
at a channel radius.  
The applicability of the approximation was tested for 
$^{16}$O+$\alpha$($^{20}$Ne) and 
$\alpha$+$\alpha$($^8$Be) systems.
It was found that the 
approximated RWA for the cluster states near the threshold energy 
is in good agreement with the exact RWA in the outer region. 
Using the approximated RWA, we estimated the $\alpha$-decay width 
in the bound state approximation
and showed that the method is useful to discuss 
the $\alpha$-decay width of resonance states. 

We applied the present method to $^9$Li, and estimate 
the partial decay width of the $^6$He($0^+_1$)+$t$ channel
for the cluster resonance states near the threshold energy. 
The present result suggests
the significant $^6$He($0^+_1$)+$t$ component in
$^9$Li($1/2^-_2$) and $^9$Li($3/2^-_2$)  at 
1$\sim$2 MeV above the threshold with the $t$-decay width of 
the order 1 MeV. 

In the present work, we apply the present method to systems consisting of simple
cluster wave functions given by H.O. configurations. The proposed method is based on the 
norm overlap with a cluster wave function localized around a certain distance 
$S_k$ which can be rather easily calculated than the exact inter-cluster wave function.
Therefore, the present method is efficient and it is applicable to systems 
consisting of more complicated cluster wave functions.
For instance, it may be feasible to evaluate the
$\alpha$ decay width of the $^{10}$Be+$\alpha$-cluster states, which has been
 theoretically suggested in excited states of $^{14}$C \cite{Suhara:2010ww}.
Moreover, application to heavier mass nuclei is promising for 
systematic study of $\alpha$-cluster states in a wide mass number region.

\section*{Acknowledgments} 
The authors would like to thank Dr.~Ogata and Dr.~Fukui for fruitful discussions.
The computational calculations of this work were performed by using the
supercomputers at YITP.
This work was supported by 
JSPS KAKENHI Grant Numbers 22540275, 25887049, 25800124, 26400270.


\begin{thebibliography}{9}


\bibitem{Ohkubo-rev}
S. Ohkubo {\it et al.}, Prog. Theor. Phys. Suppl. {\bf 132}, 1 (1998).

\bibitem{Oertzen-rev}
W. von Oertzen, M. Freer and Y. Kanada-En'yo, Phys. Rep. {\bf 432}, 43 (2006).

\bibitem{AMDsupp-rev} 
Y. Kanada-En'yo and H. Horiuchi,
Prog. Theor. Phys. Suppl. {\bf 142},  205 (2001);
Y. Kanada-En'yo M. Kimura and H. Horiuchi, 
C. R. Physique {\bf 4}, 497 (2003);
  Y.~Kanada-En'yo, M.~Kimura and A.~Ono,
  PTEP {\bf 2012} (2012) 01A202.
  
\bibitem{Horiuchi-rev}
 H. Horiuchi, K. Ikeda, and K. Kat\=o 
"Recent Developments in Nuclear Cluster Physics"
Prog. Theor. Phys. Suppl. {\bf 192}, 1 (2012).

\bibitem{Ikeda68}
K.~Ikeda, N.~Tagikawa, and H.~Horiuchi, Prog. Theor. Phys. Suppl. extra number, 464 (1968).

\bibitem{Ikeda72-supp}
K.~Ikeda {\it et al.},  Prog. Theor. Phys. Suppl. {\bf 52}, 1 (1972).


\bibitem{SEYA}
 M. Seya, M. Kohno, and S. Nagata, Prog. Theor. Phys.
 {\bf 65}, 204 (1981).
\bibitem{OERTZEN}
W. von Oertzen, Z. Phys. A {\bf 354}, 37 (1996); {\bf 357}, 355 (1997).
\bibitem{OERTZENa}
W. von Oertzen, Nuovo Cimento {\bf 110}, 895 (1997).
\bibitem{ARAI}
K. Arai, Y. Ogawa, Y. Suzuki and K. Varga, Phys. Rev. C {\bf 54}, 132 (1996).
\bibitem{Dote:1997zz}
  A.~Dote, H.~Horiuchi and Y.~Kanada-En'yo,
  Phys.\ Rev.\  C {\bf 56}, 1844 (1997).
\bibitem{Fujimura:1999zz}
  K.~Fujimura, D.~Baye, P.~Descouvemont, Y.~Suzuki and K.~Varga,
  Phys.\ Rev.\  C {\bf 59}, 817 (1999).
\bibitem{ENYObe10}
Y. Kanada-En'yo, H. Horiuchi and A. Dot\'e,
Phys. Rev. C {\bf 60}, 064304 (1999).
\bibitem{ITAGAKI}
N. Itagaki and S. Okabe, Phys. Rev. C {\bf 61}, 044306 (2000);
N. Itagaki, S. Okabe and K. Ikeda, Phys. Rev. C {\bf 62}, 034301 (2000).
\bibitem{OGAWA}
Y.Ogawa, K.Arai, Y.Suzuki and K.Varga, Nucl. Phys. {\bf A673}, 122 (2000).
\bibitem{Arai01}
K. Arai, Y. Ogawa, Y. Suzuki, and K. Varga, Prog. Theor. Phys. Suppl. 
{\bf 142}, 97 (2001).
\bibitem{Descouvemont02}
 P.~Descouvemont, Nucl. Phys. A {\bf 699}, 463 (2002).

\bibitem{KanadaEn'yo:2002ay}
  Y.~Kanada-En'yo,
  Phys.\ Rev.\  C {\bf 66}, 011303 (2002).

\bibitem{KanadaEn'yo:2003ue}
  Y.~Kanada-En'yo and H.~Horiuchi,
  Phys.\ Rev.\  C {\bf 68}, 014319 (2003).
\bibitem{Ito:2003px}
  M.~Ito, K.~Kato and K.~Ikeda,
  Phys.\ Lett.\  B {\bf 588}, 43 (2004).
\bibitem{Arai:2004yf}
  K.~Arai,
  Phys.\ Rev.\  C {\bf 69}, 014309 (2004).
\bibitem{Ito:2005yy}
  M.~Ito,
  Phys.\ Lett.\  B {\bf 636}, 293 (2006).
\bibitem{Pei:2006xg} 
  J.~C.~Pei and F.~R.~Xu,
  Phys.\ Lett.\ B {\bf 650}, 224 (2007)
  [nucl-th/0612025].
 \bibitem{Ito:2008zza}
  M.~Ito, N.~Itagaki, H.~Sakurai and K.~Ikeda,
  Phys.\ Rev.\ Lett.\  {\bf 100}, 182502 (2008).



\bibitem{Soic96}
N. Soi\'c {\it et al.}, Europhys. Lett. {\bf 34}, 7 (1996).

\bibitem{FREER}
M. Freer, {\it et al.}, Phys. Rev. Lett. {\bf 82}, 1383 (1999);
M. Freer, {\it et al.}, Phys. Rev. C {\bf 63}, 034301 (2001).

\bibitem{Liendo:2002gx}
  J.~A.~Liendo, N.~Curtis, D.~D.~Caussyn, N.~R.~Fletcher and T.~Kurtukian-Nieto,
  Phys.\ Rev.\  C {\bf 65}, 034317 (2002).


\bibitem{SAITO04}
A. Saito, {\it et al.}, Nucl. Phys. {\bf A738}, 337 (2004).
\bibitem{Curtis:2004wr}
  N.~Curtis {\it et al.},
  Phys.\ Rev.\  C {\bf 70}, 014305 (2004).
  
\bibitem{Millin05}
   M. Milin {\it et al.}, Nucl. Phys. {\bf A753}, 263 (2005).


\bibitem{Freer:2006zz}
  M.~Freer {\it et al.},
  Phys.\ Rev.\ Lett.\  {\bf 96}, 042501 (2006).


 
\bibitem{Bohlen:2007qx}
  H.~G.~Bohlen, T.~Dorsch, T.~Kokalova, W.~von Oertzen, C.~Schulz and C.~Wheldon,
  Phys.\ Rev.\  C {\bf 75}, 054604 (2007).


\bibitem{Curtis:2009zz} 
  N.~Curtis, N.~I.~Ashwood, M.~Freer, T.~Munoz-Britton, C.~Wheldon, V.~A.~Ziman, S.~Brown and W.~N.~Catford {\it et al.},
  J.\ Phys.\ G {\bf 36}, 015108 (2009).


\bibitem{Soic:2003yg} 
  N.~Soic, M.~Freer, L.~Donadille, N.~M.~Clarke, P.~J.~Leask, W.~N.~Catford, K.~L.~Jones and D.~Mahboub {\it et al.},
  Phys.\ Rev.\ C {\bf 68}, 014321 (2003).

\bibitem{oertzen04}
W. von Oertzen {\it et al.}, Eur. Phys. J. {\bf A 21}, 193 (2004).    
  
\bibitem{Price:2007mm} 
  D.~L.~Price, M.~Freer, N.~I.~Ashwood, N.~M.~Clarke, N.~Curtis, L.~Giot, V.~Lima and P.~M.~Ewan {\it et al.},
  Phys.\ Rev.\ C {\bf 75}, 014305 (2007).
  
\bibitem{Haigh:2008zz} 
  P.~J.~Haigh, N.~I.~Ashwood, T.~Bloxham, N.~Curtis, M.~Freer, P.~McEwan, D.~Price and V.~Ziman {\it et al.},
  Phys.\ Rev.\ C {\bf 78}, 014319 (2008).

\bibitem{Suhara:2010ww} 
  T.~Suhara and Y.~Kanada-En'yo,
  Phys.\ Rev.\ C {\bf 82}, 044301 (2010).



\bibitem{Descouvemont:1985zz} 
  P.~Descouvemont and D.~Baye,
  Phys.\ Rev.\ C {\bf 31}, 2274 (1985).
  
\bibitem{Gai:1983zz} 
  M.~Gai, M.~Ruscev, A.~C.~Hayes, J.~F.~Ennis, R.~Keddy, E.~C.~Schloemer, S.~M.~Sterbenz and D.~A.~Bromley,
  Phys.\ Rev.\ Lett.\  {\bf 50}, 239 (1983).
  
\bibitem{Gai:1987zz} 
  M.~Gai, R.~Keddy, D.~A.~Bromley, J.~W.~Olness and E.~K.~Warburton,
  Phys.\ Rev.\ C {\bf 36}, 1256 (1987).

\bibitem{Furutachi:2007vz} 
  N.~Furutachi, S.~Oryu, M.~Kimura, A.~Dote and Y.~Kanada-En'yo,
  Prog.\ Theor.\ Phys.\  {\bf 119}, 403 (2008).
\bibitem{Fu:2008zzf} 
  C.~Fu, V.~Z.~Goldberg, G.~V.~Rogachev, G.~Tabacaru, G.~G.~Chubarian, B.~Skorodumov, M.~McCleskey and Y.~Zhai {\it et al.},
  Phys.\ Rev.\ C {\bf 77}, 064314 (2008).  
  
\bibitem{Johnson:2009kj} 
  E.~D.~Johnson, G.~V.~Rogachev, V.~Z.~Goldberg, S.~Brown, D.~Robson, A.~M.~Crisp, P.~D.~Cottle and C.~Fu {\it et al.},
  Eur.\ Phys.\ J.\ A {\bf 42}, 135 (2009).
\bibitem{oertzen-o18}
W. von Oertzen {\it et al.},  Eur.~Phys.~J. {\bf A 43}, 17 (2010).

\bibitem{Curtis:2002mg} 
  N.~Curtis, D.~D.~Caussyn, C.~Chandler, M.~W.~Cooper, N.~R.~Fletcher, R.~W.~Laird and J.~Pavan,
  Phys.\ Rev.\ C {\bf 66}, 024315 (2002).
  
\bibitem{Ashwood:2006sb} 
  N.~I.~Ashwood, M.~Freer, S.~Ahmed, N.~M.~Clarke, N.~Curtis, P.~McEwan, C.~J.~Metelko and V.~Ziman {\it et al.},
  J.\ Phys.\ G {\bf 32}, 463 (2006).

\bibitem{Yildiz:2006xc} 
  S.~Yildiz, M.~Freer, N.~Soic, S.~Ahmed, N.~I.~Ashwood, N.~M.~Clarke, N.~Curtis and B.~R.~Fulton {\it et al.},
  Phys.\ Rev.\ C {\bf 73}, 034601 (2006).
\bibitem{Scholz:1972zz} 
  W.~Scholz, P.~Neogy, K.~Bethge and R.~Middleton,
  Phys.\ Rev.\ C {\bf 6}, 893 (1972).
  
\bibitem{Descouvemont:1988zz} 
  P.~Descouvemont,
  Phys.\ Rev.\ C {\bf 38}, 2397 (1988).
\bibitem{Kimura:2007kz} 
  M.~Kimura,
  Phys.\ Rev.\ C {\bf 75}, 034312 (2007).
  
\bibitem{Rogachev:2001ti} 
  G.~V.~Rogachev, V.~Z.~Goldberg, T.~Lonnroth, W.~H.~Trzaska, S.~A.~Fayans, K.~-M.~Kallman, J.~J.~Kolata and M.~Mutterer {\it et al.},
  Phys.\ Rev.\ C {\bf 64}, 051302 (2001).
  
\bibitem{Goldberg:2004yk} 
  V.~Z.~Goldberg, G.~V.~Rogachev, W.~H.~Trzaska, J.~J.~Kolata, A.~Andreyev, C.~Angulo, M.~J.~G.~Borge and S.~Cherubini {\it et al.},
  Phys.\ Rev.\ C {\bf 69}, 024602 (2004).

\bibitem{KanadaEn'yo:2011nc} 
  Y.~Kanada-En'yo and T.~Suhara,
  Phys.\ Rev.\ C {\bf 85}, 024303 (2012).

\bibitem{RGM}
J. A. Wheeler, Phys. Rev. {\bf 52} 1083 (1937);
J. A. Wheeler, Phys. Rev. {\bf 52} 1107 (1937).

\bibitem{wildermuth58}
K. Wildermuth, Th. Kanellopoulos,
Nucl. Phys., {\bf 7}, 150 (1958);
K. Wildermuth, Th. Kanellopoulos,
Nucl. Phys., {\bf 9}, 449 (1958/1959).

\bibitem{GCM}
 D. L. Hill and J. A. Wheeler, Phys. Rev. {\bf 89}, 1102 (1953);
J. J. Griffin and J. A. Wheeler, Phys. Rev. {\bf 108}, 311 (1957). 

\bibitem{brink66} D. M. Brink, International School of Physics ``Enrico Fermi'', XXXVI, p. 247 (1966).

\bibitem{tamagaki65}
R. Tamagaki, H. Tanaka, Prog. Theor. Phys. {\bf 34}, 191 (1965);
R. Tamagaki, Prog. Theor. Phys. Suppl. Extra Number, 242 (1968);
J. Hiura and R. Tamagaki, Prog. Theor. Phys. Suppl. {\bf 52}, 25 (1972).


\bibitem{wildermuth72}
W. S\"unkel, K. Mildermuth, Phys. Lett. {\bf B41}, 439 (1972). 

\bibitem{Nemoto72}
F. Nemoto and H. Band\=o,
Prog. Theor. Phys. {\bf 47}, 1210 (1971).

\bibitem{Matsuse73}
T. Matsuse, M. Kamimura, and Y. Fukushima, 
Prog. Theor. Phys. {\bf 49}, 1765 (1973).

\bibitem{Matsuse75}
T. Matsuse, M. Kamimura, and Y. Fukushima, 
Prog. Theor. Phys. {\bf 53}, 706 (1975).

\bibitem{Ikeda77-supp}
K. Ikeda {\it et al.}, Prog. Theor. Phys. Suppl. {\bf 62}, 1 (1977).

\bibitem{KanadaEnyo:1994kw} 
  Y.~Kanada-Enyo and H.~Horiuchi,
  Prog.\ Theor.\ Phys.\  {\bf 93}, 115 (1995).

\bibitem{ENYObc}
  Y.~Kanada-En'yo, H.~Horiuchi and A.~Ono,
  Phys.\ Rev.\  C {\bf 52}, 628  (1995);
  Y.~Kanada-En'yo and H.~Horiuchi,
  Phys.\ Rev.\  C {\bf 52}, 647 (1995).

\bibitem{KanadaEn'yo:1998rf} 
  Y.~Kanada-En'yo,
  Phys.\ Rev.\ Lett.\  {\bf 81}, 5291 (1998).


\bibitem{Feldmeier:1994he} 
  H.~Feldmeier, K.~Bieler and J.~Schnack,
  Nucl.\ Phys.\ A {\bf 586}, 493 (1995).
\bibitem{Feldmeier:2000cn} 
  H.~Feldmeier and J.~Schnack,
  Rev.\ Mod.\ Phys.\  {\bf 72}, 655 (2000).

\bibitem{Roth:2004ua} 
  R.~Roth, T.~Neff, H.~Hergert and H.~Feldmeier,
  Nucl.\ Phys.\ A {\bf 745}, 3 (2004).

\bibitem{Neff:2010nm} 
  T.~Neff,
  Phys.\ Rev.\ Lett.\  {\bf 106}, 042502 (2011).
  
\bibitem{Varga:1993wp} 
  K.~Varga, Y.~Suzuki and R.~G.~Lovas,
  Nucl.\ Phys.\ A {\bf 571}, 447 (1994).
  
\bibitem{Kimura:2003ue} 
  M.~Kimura and H.~Horiuchi,
  Phys.\ Rev.\ C {\bf 69}, 051304 (2004).





\bibitem{VOLKOV} A. B. Volkov, Nucl. Phys. {\bf 74}, 33 (1965).



\bibitem{LS}
 N. Yamaguchi, T. Kasahara, S. Nagata and Y. Akaishi,
 {\em Prog. Theor. Phys.} {\bf 62}, 1018  (1979);
 R. Tamagaki, {\em Prog. Theor. Phys.} {\bf 39}, 91  (1968).
 
\bibitem{Tilley98}
	D. R. Tilley, C. Cheves, J. Kelley, S. Raman, H. Weller, 
  Nucl.\ Phys.\ A {\bf 636}, 249 (1993).
  
\bibitem{Kelley04}
J.~H.~Kelley, J.~L.=Godwin, C.~G.~Sheu, {\it et al.},
  Nucl.\ Phys.\ A {\bf 745}, 155 (2004). 


\end{thebibliography}

\end{document}